\documentclass[usenatbib]{mnras}
\pdfoutput=1

\usepackage[fleqn]{amsmath}
\usepackage{amssymb}
\usepackage{amsfonts}
\usepackage{lineno}

\usepackage{definitions}

\usepackage{graphicx}
\graphicspath{{./pdf/}}

\usepackage{hyperref}

\def\percent{\%}

\title[Reliable uncertainties in IR interferometry]{
    Towards reliable uncertainties in IR interferometry:\\
    The bootstrap for correlated statistical \& systematic errors}
\author[Lachaume et al.]{%
R\'egis Lachaume,$^{1,2}$\thanks{regis.lachaume@gmail.com}
Markus Rabus,$^{1,2,3,4}$ 
Andr\'es Jord\'an,$^{1,5}$
\newauthor 
Rafael Brahm,$^{1,5}$ 
Tabetha Boyajian,$^{6}$
Kaspar von Braun,$^{7}$ 
Jean-Philippe Berger$^{8}$ 
\\
$^{1}$Instituto de Astronom\'\i{}a, Facultad de F\'\i{}sica, Pontificia Universidad Cat\'olica de Chile, casilla 306, Santiago 22, Chile\\
$^{2}$Max-Planck-Institut f\"ur Astronomie, K\"onigstuhl 17, D-69117 Heidelberg, Germany\\
$^{3}$Las Cumbres Observatory Global Telescope, 6740 Cortona Dr., Suite 102, Goleta, CA 93111, USA\\
$^{4}$Department of Physics, University of California, Santa Barbara, CA 93106-9530, USA\\
$^{5}$Millennium Institute of Astrophysics\\
$^{6}$Department of Physics and Astronomy, Louisiana State University, 202 Nicholsom Hall, Baton Rouge, LA 70803, USA\\
$^{7}$Lowell Observatory, 1400 W. Mars Hill Road, Flagstaff, AZ 86001, USA\\ 
$^{8}$Institut de Plan\'etologie et d'Astrophysique de Grenoble, Observatoire de Grenoble, France
}

\def\change#1{#1}

\begin{document}

\maketitle

\begin{abstract}
We propose a method to overcome the usual limitation of current data processing techniques in optical and infrared long-baseline interferometry: most reduction pipelines assume uncorrelated statistical errors and ignore systematics. We use the bootstrap method to sample the multivariate probability density function of the interferometric observables. It allows us to determine the correlations between statistical error terms and their deviation from a Gaussian distribution. In addition, we introduce systematics as an additional, highly correlated error term whose magnitude is chosen to fit the data dispersion. 

We have applied the method to obtain accurate measurements of stellar diameters for under-resolved stars, i.e. smaller than the angular resolution of the interferometer. We show that taking correlations and systematics has a significant impact on both the diameter estimate and its uncertainty.  The robustness of our diameter determination comes at a price: we obtain 4 times larger uncertainties, of a few \change{percent} for most stars in our sample.
\end{abstract}


\section{Introduction}

Long-baseline interferometry consists in recombining the light from several telescopes to measure interference fringes on an astronomical object \citep{LAW00}.  The key observable is the visibility, a complex number containing the contrast and phase of fringes obtained on a telescope pair. It was first introduced in radioastronomy \citep{BRA58} to generalise Michelson's term \citep[a synonym for contrast in][]{MIC21}.  In the ideal case, the visibility is the Fourier transform of the object's image taken at a spatial frequency related to the telescope separation \citep{BRA58,HON75,LAB75}.  In the infrared and optical, though, the atmospheric turbulence shifts the fringes in milliseconds, so that only the square visibility amplitude and partial phase information can be retrieved \citep{ROD84b}, for instance via the closure phase, the sum of phases over a telescope triplet \citep{ROD86,COR87}. A robust determination of the uncertainties on these observables is paramount to ensure confidence on the physical parameters derived from model fitting. 

\textit{``Statistical'' errors}---deviations of expected mean zero---on interferometric observables are produced by fast-varying intrinsic, atmospheric, and instrumental effects.  In addition to the detector and photon noises, a few sources of errors that have been observed at the Very Large Telescope Interferometer (VLTI) are the differential atmospheric piston \citep{COL99,ESP00}, imperfect fibre injection \citep{KOT03}, mechanical vibrations \citep[][Sect.~5.6]{PIONIER}, background fluctuations \citep{ABS04}, detector efficiency variations due to cooling cycle \citep{ABS04}, and 50\,Hz electronic noise from the power grid \citep{ABS04}. The relatively large number of sources of uncertainties made it difficult to obtain a reliable assessment of the precision of interferometric measurements. In particular, the theoretical estimate using photon and detector noises is unrealistically low, so most processing software tools need heuristics to provide a better one, for instance using the dispersion of a given data set. One of the common pitfalls is the assumption that the statistical errors on visibility measurements are uncorrelated \citep{MEI05} and follow a Gaussian distribution. In particular, most public data processing software tools do not determine these correlations, in particular those for the VLTI instruments MIDI\footnote{MID-infrared Interferometric instrument} \citep{MIDI}, AMBER\footnote{Astronomical Multi-BEam combineR} \citep{AMBER}, PIONIER\footnote{Precision Integrated-Optics Near-infrared Imaging ExpeRiment} \citep{PIONIER}, and GRAVITY \citep{GRAVITYpipe}.   The same happens with popular model-fitting tools \citep[e.g. Litpro, see][]{TAL08} or image reconstruction programmes \citep[e.g. MIRA, see][]{THI08}.  Also, the Optical Interferometric FITS \citep[OIFITS v. 1, see][]{OIFITS1} format did not provide a codified way to document correlations in its first and most used version. Only very recently has an update to the standard given specifications for a covariance matrix \citep[OIFITS v. 2, see][]{OIFITS2}.

The assumption of uncorrelated Gaussian measurement errors could not be further from the truth. For instance, closure phases are not independent \citep{MON07}. Also, several random effects impact the different spectral channels of a same observation in the same way, such as the blurring of fringes due to the turbulent atmosphere \citep[][Sect.~7.5 ``Atmospheric Biases'']{LAW00}. Finally, all observations of an observing sequence are impacted in the same way by the errors on the calibrators, virtually leading to correlations between all data points \citep{PER03}, even collected in different runs at different facilities. The assumption that the fringe contrasts or phases follow a Gaussian distribution is not confirmed by experience either \citep[][in the case of AMBER]{SCH14}.  Also, when deriving the instrumental transfer function, a weighted average of square visibilities and closure/differential phases is obtained, using a few calibrators observed close to the science targets.  In most cases this average does not follow Gaussian distribution, as \citet{PER03} note.   

For these reasons, \citet{PER03} proposed an analytic formalism to propagate the non-Gaussian correlated uncertainties of the square visibility amplitudes in an approximate, yet relatively accurate way.  Several authors have applied these results to the FLUOR\footnote{Fiber Linked Unit for Optical Recombination} instrument \citep[at IOTA\footnote{Infrared and Optical Telescope Array}, then CHARA\footnote{Center for High Angular Resolution Array}, see for instance][]{PER04,ABS06,BER06} to our knowledge the only one for which correlations have been regularly determined.  

In addition to the statistical errors that one can infer from the data and/or noise modelling, there are \emph{``systematic'' errors} that typically plague interferometric data and impact all the data of a given set of observations in a similar and poorly understood way.  Part of the biases are removed either theoretically \citep[e.g. group-delay dispersion, see][]{ZYV03} or calibrated out \citep[e.g. polarisation,][]{HAG00} by the measurement of the ``instrumental visibility'' on stars of known geometry, ideally unresolved ones, with the underlying assumption that it varies slowly enough to be interpolated to science observations with sufficient precision \citep{HAN74,PER03}.

However, complete removal does not happen. \citet{COL03} measured $\approx 5\%$ systematic errors on the calibrated squared visibility amplitudes at the Keck Interferometer by observing binaries of known orbital parameters.  More recently, high-precision diameter measurements using sufficiently well resolved stars with CHARA \citep{WHI18,KAR18} were shown to significantly differ (several $\sigma$ and up to 15\%) from values previously obtained from under-resolved interferometric observations, leading the authors to conclude that they were plagued with undiagnosed systematics. At VLTI, \citet{LEB09} and \citet{KER04} also tried to identify the origins of the large systematics, discarding the uncertainty on central wavelength \citep[calibration errors are reported to be 0.35 to 0.50\percent{} at PIONIER at VLTI and PAVO\footnote{Precision Astronomical Visible Observations} at CHARA, respectively by][]{HUB12,GAL18}, atmospheric jitter and injection efficiency due to seeing, and instrumental variations during the observation ($\kappa$ matrix). Possible sources are a differential polarisation effect \citep[several percents,][]{LEB12} and the way the bias is removed in Fourier space during the PIONIER data processing \citep[see][in the case of FINITO\footnote{Fringe-tracking Instrument of NIce and TOrino}]{LEB09}.  We also stress that calibrators may be an additional source of systematics: for instance, an unsuspected binary with a flux ratio of 1:100 would likely go undetected in closure phase with PIONIER ($\lesssim 2.3$\,deg), yet could account for a bias in the squared visibility amplitudes of up to 2\%. 

While some authors deal with the uncertainty on the calibrators' diameters as a systematic error \citep[e.g.][but we include them in the correlated ``statistical'' errors for practical reasons]{CRE15,PER16}, few take into account other systematics, despite all the evidence. \citet{HUB12} introduce systematic errors of the order of one percent in order to account for errors in the spectral calibration, but they are still much smaller than the observed and unexplained errors.

In this paper, we present a model that make use of a technique that is relatively easy to implement into existing pipelines, is more accurate than the classical error propagation, and requires no additional analytic developments. Known as the bootstrap method \citep{BOOTSTRAP}, it consists in randomly selecting interferograms to feed the data reduction software.  Repeating it enough times, we generate a sampling of the multivariate probability density function of the square visibility amplitudes and closure phases.  Bootstrapping was originally introduced in interferometry by \citet{KER04drs} in order to determine the statistical errors for VINCI\footnote{\change{VLT Interferometer commissioning instrument}} at the VLTI.  In addition, we treat the systematic errors as additional correlated term whose magnitude is left as a free parameter. 

In Sect.~\ref{sec:obs}, we present the data set, acquired for the companion paper by \citet{RAB18}, that has led us to undertake this work.  Sect.~\ref{sec:dataproc} details our modelling of the correlated statistical errors and systematic errors.  We then show (Sect.~\ref{sec:ud}) how the estimate uniform disc diameters with PIONIER at the VLTI is significantly impacted by the level of detail in the error modelling.  We summarise and conclude in Sect.~\ref{sec:conclusion}.

\section{Data set}
\label{sec:obs} 

\begin{table*}
\caption{Observing log sorted by science target, calibrator, and observing night. Calibrator characteristics are $H$ magnitude, spectral type, uniform disc diameter $\vartheta_\mathrm{cal}$, and relative weight in the transfer function calculation, with 1.0 for calibrators very close in time and space (e.g. SCI-CAL-SCI block). This weight does not include the lesser impact of large diameter, large diameter uncertainties, and large visibility dispersion under bad conditions that arises from error propagation. Sky conditions in the direction of the target in $H$ band are the seeing and the atmospheric phase coherence time $\tau_0$.  Science and calibrators observed close to, or above, the nominal limiting magnitude of the instrument are listed as $\approx$~lim and $>$~lim. We have fitted $H_\text{lim} = 6.7 - 2.5 \log_{10}(\text{seeing in $V$)}$ to the conservative estimates from the ESO Call for Proposals, albeit it is better when $\tau_0$ is large (e.g. GJ~1061).}

\label{tab:obslog}
\def\pmf{\pm\phantom{0}}
\begin{tabular}{lc lclcc c crcc}
\hline
\multicolumn{2}{c}{science target} & \multicolumn{5}{c}{calibrator} & night & \multicolumn{4}{c}{conditions} \\
\multicolumn{1}{c}{name} & $H$ & \multicolumn{1}{c}{name} & $H$ & \multicolumn{1}{c}{sp. type} & $\vartheta_\mathrm{cal}$ & weight & night & seeing & \multicolumn{1}{c}{$\tau_0$} & \multicolumn{2}{c}{ins. limit} \\
 & [mag] &  & [mag] &  & [mas] & [MJD] &  & [arcsec] & \multicolumn{1}{c}{[ms]} & sci. & cal. \\
\hline
GJ 1       & 4.73 & HD 1434    & 3.75 & K2III    & $0.941 \pm 0.013$ &  0.96 &  56596 & $0.68 \pm 0.04$ & $ 6.8 \pmf 0.4$ &        &       \\
           &      &            &      &          &                   &  0.86 &  56849 & $1.36 \pm 0.17$ & $ 3.3 \pmf 0.4$ &        &       \\
           &      &            &      &          &                   &  0.92 &  56850 & ---             & $17.8 \pmf 0.0$ &        &       \\
           &      & HD 190     & 5.38 & G8III    & $0.431 \pm 0.085$ &  0.64 &  56849 & $1.13 \pm 0.08$ & $ 3.8 \pmf 0.3$ &        &       \\
           &      &            &      &          &                   &  0.76 &  56850 & ---             & $17.8 \pmf 0.0$ &        &       \\
           &      &            &      &          &                   &  0.98 &  56889 & $0.75 \pm 0.13$ & $22.4 \pmf 3.1$ &        &       \\
           &      & HD 214623  & 4.05 & K2/3III  & $0.776 \pm 0.011$ &  0.22 &  56597 & $0.92 \pm 0.05$ & $ 5.8 \pmf 0.4$ &        &       \\
           &      & HD 215709  & 5.61 & F6V      & $0.313 \pm 0.062$ &  0.28 &  56889 & $0.57 \pm 0.05$ & $28.3 \pmf 2.6$ &        &       \\
           &      & HD 216988  & 5.17 & K0III    & $0.468 \pm 0.093$ &  0.22 &  56597 & $0.93 \pm 0.04$ & $ 5.4 \pmf 0.2$ &        &       \\
           &      & HD 217681  & 5.64 & K0III    & $0.362 \pm 0.072$ &  0.62 &  56889 & $0.58 \pm 0.05$ & $28.7 \pmf 2.2$ &        &       \\
           &      & HD 224936  & 4.66 & K1III    & $0.635 \pm 0.126$ &  0.53 &  56596 & $0.68 \pm 0.10$ & $ 6.6 \pmf 0.9$ &        &       \\
           &      &            &      &          &                   &  0.95 &  56890 & $0.81 \pm 0.12$ & $14.7 \pmf 1.9$ &        &       \\
           &      & HD 224949  & 4.92 & K0III    & $0.556 \pm 0.110$ &  0.93 &  56596 & $0.75 \pm 0.05$ & $ 6.0 \pmf 0.4$ &        &       \\
           &      &            &      &          &                   &  0.90 &  56597 & $0.91 \pm 0.05$ & $ 5.6 \pmf 0.3$ &        &       \\
           &      &            &      &          &                   &  0.84 &  56849 & $1.15 \pm 0.07$ & $ 3.7 \pmf 0.2$ &        &       \\
           &      &            &      &          &                   &  0.84 &  56850 & ---             & $17.9 \pmf 0.1$ &        &       \\
           &      &            &      &          &                   &  0.97 &  56889 & $0.85 \pm 0.17$ & $20.9 \pmf 3.5$ &        &       \\
           &      & HD 225101  & 6.23 & F2V      & $0.234 \pm 0.047$ &  0.95 &  56597 & $0.95 \pm 0.05$ & $ 5.4 \pmf 0.3$ &        &       \\
           &      & HD 32      & 4.80 & G8III/IV & $0.538 \pm 0.107$ &  0.91 &  56596 & $0.80 \pm 0.13$ & $ 5.8 \pmf 0.9$ &        &       \\
           &      &            &      &          &                   &  0.90 &  56597 & $0.86 \pm 0.07$ & $ 5.9 \pmf 0.4$ &        &       \\
           &      &            &      &          &                   &  0.89 &  56890 & $0.95 \pm 0.17$ & $12.9 \pmf 2.0$ &        &       \\
           &      & HD 902     & 3.77 & K3/4III  & $0.967 \pm 0.013$ &  0.78 &  56596 & $0.62 \pm 0.05$ & $ 7.4 \pmf 0.6$ &        &       \\
           &      &            &      &          &                   &  0.93 &  56849 & $1.51 \pm 0.09$ & $ 3.1 \pmf 0.1$ &        &       \\
           &      &            &      &          &                   &  0.88 &  56850 & ---             & $17.9 \pmf 0.0$ &        &       \\
\hline
GJ 54.1    & 6.75 & HD 22000   & 7.13 & G8III    & $0.187 \pm 0.037$ &  0.54 &  56889 & $0.97 \pm 0.22$ & $21.3 \pmf 3.8$ & $\approx$ lim & $>$ lim\\
           &      & HD 5248    & 7.28 & K0V      & $0.163 \pm 0.032$ &  0.92 &  56889 & $0.84 \pm 0.05$ & $20.2 \pmf 1.3$ & $\approx$ lim & $>$ lim\\
           &      & HD 5911    & 6.51 & G3V      & $0.216 \pm 0.043$ &  0.89 &  56597 & $1.03 \pm 0.17$ & $ 5.8 \pmf 0.7$ & $>$ lim & $\approx$ lim\\
           &      & HD 5932    & 6.99 & F0V      & $0.156 \pm 0.031$ &  0.92 &  56890 & $0.82 \pm 0.20$ & $13.6 \pmf 2.8$ & $\approx$ lim & $\approx$ lim\\
           &      & HD 6022    & 7.15 & F2V      & $0.154 \pm 0.031$ &  0.94 &  56889 & $0.87 \pm 0.03$ & $20.2 \pmf 1.1$ & $\approx$ lim & $>$ lim\\
           &      & HD 6482    & 3.81 & K0III    & $0.836 \pm 0.012$ &  0.84 &  56596 & $0.92 \pm 0.08$ & $ 6.3 \pmf 1.0$ & $\approx$ lim &       \\
           &      & HD 6720    & 6.11 & G8V      & $0.288 \pm 0.057$ &  0.84 &  56542 & ---             & $ 5.9 \pmf 0.1$ & $\approx$ lim &       \\
           &      &            &      &          &                   &  0.96 &  56596 & $0.98 \pm 0.12$ & $ 4.9 \pmf 0.6$ & $\approx$ lim &       \\
           &      & HD 7257    & 6.58 & F3V      & $0.210 \pm 0.042$ &  0.84 &  56542 & ---             & $ 6.3 \pmf 0.3$ & $\approx$ lim & $\approx$ lim\\
           &      &            &      &          &                   &  0.86 &  56597 & $0.86 \pm 0.14$ & $ 6.1 \pmf 0.8$ & $\approx$ lim &       \\
           &      & HD 7495    & 6.26 & F6V      & $0.235 \pm 0.047$ &  0.97 &  56596 & $0.71 \pm 0.06$ & $ 7.4 \pmf 0.5$ &        &       \\
           &      &            &      &          &                   &  0.76 &  56597 & $0.90 \pm 0.05$ & $ 5.4 \pmf 0.3$ & $\approx$ lim &       \\
           &      & HD 8406    & 6.50 & G3V      & $0.232 \pm 0.046$ &  0.73 &  56596 & $0.63 \pm 0.04$ & $ 6.9 \pmf 0.4$ &        &       \\
           &      &            &      &          &                   &  0.91 &  56597 & $0.97 \pm 0.05$ & $ 6.6 \pmf 0.8$ & $\approx$ lim & $\approx$ lim\\
\hline
GJ 86      & 4.25 & HD 10939   & 5.03 & A1V      & $0.329 \pm 0.066$ &  0.98 &  56253 & $0.63 \pm 0.03$ & $10.6 \pmf 0.4$ &        &       \\
           &      & HD 12619   & 3.52 & K5III    & $1.139 \pm 0.016$ &  0.85 &  56210 & $0.68 \pm 0.14$ & $ 7.9 \pmf 1.4$ &        &       \\
           &      &            &      &          &                   &  0.98 &  56252 & $0.84 \pm 0.22$ & $ 6.8 \pmf 1.8$ &        &       \\
           &      & HD 13666   & 3.21 & K2/3III  & $1.007 \pm 0.014$ &  0.95 &  56252 & $0.84 \pm 0.23$ & $ 6.5 \pmf 1.9$ &        &       \\
           &      & HD 15371   & 4.66 & B7IV     & $0.402 \pm 0.080$ &  0.98 &  56253 & $0.70 \pm 0.05$ & $ 8.9 \pmf 0.6$ &        &       \\
           &      & HD 15520   & 4.61 & K1III    & $0.624 \pm 0.124$ &  0.85 &  56210 & $0.74 \pm 0.12$ & $ 7.4 \pmf 1.3$ &        &       \\
           &      & HD 20640   & 3.09 & K2III    & $1.260 \pm 0.016$ &  0.92 &  56252 & $0.92 \pm 0.10$ & $ 5.9 \pmf 0.6$ &        &       \\
           &      & HD 21011   & 3.93 & K0III    & $0.688 \pm 0.009$ &  0.52 &  56253 & $0.88 \pm 0.21$ & $ 7.3 \pmf 1.8$ &        &       \\
           &      & HD 25038   & 4.29 & K2III    & $0.812 \pm 0.011$ &  0.27 &  56253 & $0.61 \pm 0.04$ & $ 9.3 \pmf 0.6$ &        &       \\
           &      & HD 26934   & 3.60 & K3III    & $0.984 \pm 0.013$ &  0.66 &  56253 & $0.74 \pm 0.12$ & $ 8.0 \pmf 1.2$ &        &       \\
           &      & HD 22663  & 2.14 & K1III    & $1.890 \pm 0.022$ &  0.68 &  56252 & $1.45 \pm 0.44$ & $ 3.8 \pmf 1.1$ &        &       \\
           &      & HD 28776   & 3.63 & K0II     & $0.837 \pm 0.166$ &  0.90 &  56253 & $0.57 \pm 0.10$ & $10.9 \pmf 1.7$ &        &       \\
\hline
\end{tabular}
\end{table*}
\begin{table*}
\contcaption{Observing log}
\def\pmf{\pm\phantom{0}}
\begin{tabular}{lc lclcc c crcc}
\hline
\multicolumn{2}{c}{science target} & \multicolumn{5}{c}{calibrator} & night & \multicolumn{4}{c}{conditions} \\
\multicolumn{1}{c}{name} & $H$ & \multicolumn{1}{c}{name} & $H$ & \multicolumn{1}{c}{sp. type} & $\vartheta_\mathrm{cal}$ & weight & night & seeing & \multicolumn{1}{c}{$\tau_0$} & \multicolumn{2}{c}{ins. limit} \\
 & [mag] &  & [mag] &  & [mas] & [MJD] &  & [arcsec] & \multicolumn{1}{c}{[ms]} & sci. & cal. \\
\hline
GJ 229     & 4.39 & HD 38090   & 5.54 & A2/3V    & $0.272 \pm 0.054$ &  0.98 &  56325 & $1.32 \pm 0.27$ & $ 5.2 \pmf 0.9$ &        &       \\
           &      & HD 40379   & 5.49 & F6V      & $0.330 \pm 0.066$ &  0.74 &  56325 & $1.21 \pm 0.17$ & $ 4.4 \pmf 0.5$ &        &       \\
           &      & HD 43445   & 5.12 & B9V      & $0.275 \pm 0.055$ &  0.76 &  56253 & $0.50 \pm 0.03$ & $10.9 \pmf 0.7$ &        &       \\
           &      &            &      &          &                   &  0.88 &  56325 & $1.29 \pm 0.25$ & $ 5.3 \pmf 1.3$ &        &       \\
           &      & HD 44893   & 4.22 & K2/3III  & $0.800 \pm 0.010$ &  0.74 &  56253 & $0.53 \pm 0.10$ & $10.4 \pmf 2.2$ &        &       \\
           &      & HD 48286   & 5.58 & F9V      & $0.329 \pm 0.066$ &  0.97 &  56325 & $1.46 \pm 0.17$ & $ 4.0 \pmf 0.2$ &        & $\approx$ lim\\
           &      & HD 58187   & 5.18 & A5IV     & $0.356 \pm 0.071$ &  0.74 &  56325 & $2.02 \pm 0.10$ & $ 3.8 \pmf 0.2$ &        & $\approx$ lim\\
           &      & HD 60111   & 4.97 & F0/2IV/V & $0.407 \pm 0.081$ &  0.36 &  56325 & $1.47 \pm 0.13$ & $ 4.4 \pmf 0.4$ &        &       \\
\hline
GJ 273     & 5.22 & HD 48286   & 5.58 & F9V      & $0.329 \pm 0.066$ &  0.37 &  56325 & $1.61 \pm 0.10$ & $ 4.0 \pmf 0.2$ &        & $\approx$ lim\\
           &      & HD 58187   & 5.18 & A5IV     & $0.356 \pm 0.071$ &  0.96 &  56325 & $2.02 \pm 0.10$ & $ 3.8 \pmf 0.2$ & $>$ lim & $\approx$ lim\\
           &      &            &      &          &                   &  0.88 &  56384 & $0.73 \pm 0.06$ & $ 8.4 \pmf 0.6$ &        &       \\
           &      &            &      &          &                   &  0.94 &  56737 & $1.34 \pm 0.08$ & $ 6.9 \pmf 0.4$ &        &       \\
           &      &            &      &          &                   &  0.93 &  56740 & $0.66 \pm 0.03$ & $11.0 \pmf 0.6$ &        &       \\
           &      & HD 58556   & 5.73 & G1V      & $0.305 \pm 0.061$ &  0.98 &  56325 & $1.24 \pm 0.09$ & $ 6.6 \pmf 0.6$ &        &       \\
           &      &            &      &          &                   &  0.72 &  56384 & $0.73 \pm 0.12$ & $ 7.2 \pmf 0.8$ &        &       \\
           &      &            &      &          &                   &  0.86 &  56737 & $0.93 \pm 0.15$ & $ 8.3 \pmf 2.4$ &        &       \\
           &      &            &      &          &                   &  0.84 &  56740 & $0.64 \pm 0.05$ & $ 8.8 \pmf 0.7$ &        &       \\
           &      & HD 60111   & 4.97 & F0/2IV/V & $0.407 \pm 0.081$ &  0.98 &  56325 & $1.47 \pm 0.13$ & $ 4.4 \pmf 0.4$ &        &       \\
           &      &            &      &          &                   &  0.92 &  56737 & $1.24 \pm 0.11$ & $ 6.0 \pmf 0.6$ &        &       \\
           &      &            &      &          &                   &  0.91 &  56740 & $0.62 \pm 0.05$ & $ 9.7 \pmf 0.7$ &        &       \\
           &      & HD 60275   & 6.19 & A0V      & $0.182 \pm 0.036$ &  0.87 &  56737 & $1.18 \pm 0.05$ & $ 8.1 \pmf 0.4$ &        & $\approx$ lim\\
           &      & HD 64685   & 5.01 & F3V      & $0.391 \pm 0.078$ &  0.96 &  56325 & $1.46 \pm 0.10$ & $ 5.8 \pmf 0.7$ &        &       \\
\hline
GJ 370     & 5.00 & HD 85483   & 3.53 & K0III    & $0.974 \pm 0.013$ &  0.81 &  56737 & $1.21 \pm 0.15$ & $ 5.7 \pmf 0.7$ &        &       \\
           &      &            &      &          &                   &  0.87 &  56738 & $0.91 \pm 0.08$ & $ 7.3 \pmf 0.5$ &        &       \\
           &      &            &      &          &                   &  0.91 &  56739 & $1.00 \pm 0.09$ & $ 5.5 \pmf 0.6$ &        &       \\
           &      & HD 85849   & 5.95 & K1III    & $0.318 \pm 0.063$ &  0.88 &  56738 & $0.89 \pm 0.08$ & $ 6.6 \pmf 0.8$ &        &       \\
           &      &            &      &          &                   &  0.93 &  56739 & $1.06 \pm 0.09$ & $ 4.9 \pmf 0.4$ &        &       \\
           &      &            &      &          &                   &  0.90 &  56740 & $0.55 \pm 0.03$ & $10.4 \pmf 0.6$ &        &       \\
           &      & HD 85996   & 5.71 & K0III    & $0.367 \pm 0.073$ &  0.96 &  56738 & $1.08 \pm 0.10$ & $ 5.1 \pmf 0.5$ &        &       \\
           &      &            &      &          &                   &  0.84 &  56739 & $0.83 \pm 0.13$ & $ 6.4 \pmf 0.9$ &        &       \\
           &      &            &      &          &                   &  0.75 &  56740 & $0.49 \pm 0.03$ & $11.5 \pmf 0.7$ &        &       \\
\hline
GJ 406     & 6.48 & HD 93081   & 6.02 & F6V      & $0.275 \pm 0.055$ &  0.68 &  56737 & $1.09 \pm 0.26$ & $ 6.7 \pmf 1.4$ & $\approx$ lim &       \\
           &      &            &      &          &                   &  0.80 &  56738 & $1.02 \pm 0.18$ & $ 5.8 \pmf 0.9$ & $\approx$ lim &       \\
           &      &            &      &          &                   &  0.88 &  56739 & $0.70 \pm 0.06$ & $ 8.3 \pmf 0.5$ &        &       \\
           &      &            &      &          &                   &  0.91 &  56740 & $0.50 \pm 0.03$ & $12.6 \pmf 0.8$ &        &       \\
           &      & HD 95216   & 5.45 & F5V      & $0.337 \pm 0.067$ &  0.54 &  56738 & $0.75 \pm 0.05$ & $ 9.3 \pmf 0.6$ &        &       \\
           &      &            &      &          &                   &  0.90 &  56740 & $0.70 \pm 0.13$ & $10.9 \pmf 1.8$ &        &       \\
\hline
GJ 433     & 5.86 & HD 103868  & 5.96 & F3V      & $0.253 \pm 0.050$ &  0.99 &  56326 & $1.39 \pm 0.02$ & $ 9.7 \pmf 1.2$ & $\approx$ lim & $\approx$ lim\\
           &      & HD 96557   & 5.72 & F2V      & $0.276 \pm 0.055$ &  0.98 &  56326 & $1.51 \pm 0.19$ & $ 7.7 \pmf 1.1$ & $>$ lim & $\approx$ lim\\
           &      &            &      &          &                   &  0.58 &  56384 & $0.59 \pm 0.11$ & $ 7.6 \pmf 1.4$ &        &       \\
           &      & HD 98220   & 5.64 & F7V      & $0.319 \pm 0.063$ &  0.97 &  56326 & $1.56 \pm 0.20$ & $ 6.9 \pmf 0.8$ & $>$ lim & $\approx$ lim\\
           &      &            &      &          &                   &  0.83 &  56384 & $0.58 \pm 0.05$ & $ 7.7 \pmf 0.6$ &        &       \\
\hline
GJ 447     & 5.95 & HD 101730  & 5.88 & F5V      & $0.294 \pm 0.059$ &  0.99 &  56325 & $0.83 \pm 0.04$ & $ 9.7 \pmf 0.9$ &        &       \\
           &      &            &      &          &                   &  0.64 &  56384 & $0.61 \pm 0.06$ & $ 8.4 \pmf 0.6$ &        &       \\
           &      & HD 103773  & 5.68 & F7V      & $0.306 \pm 0.061$ &  0.98 &  56325 & $0.82 \pm 0.06$ & $ 8.6 \pmf 0.6$ &        &       \\
           &      &            &      &          &                   &  0.87 &  56384 & $0.56 \pm 0.05$ & $ 8.4 \pmf 1.0$ &        &       \\
           &      & HD 97937   & 5.96 & F0V      & $0.251 \pm 0.050$ &  0.98 &  56325 & $1.02 \pm 0.08$ & $ 8.0 \pmf 0.8$ &        &       \\
\hline
GJ 551     & 4.84 & HD 119073  & 5.48 & K0III    & $0.398 \pm 0.079$ &  0.63 &  56737 & $0.80 \pm 0.06$ & $ 8.6 \pmf 0.6$ &        &       \\
           &      & HD 128398  & 5.38 & F7IV     & $0.342 \pm 0.068$ &  0.85 &  56849 & $1.00 \pm 0.05$ & $ 4.9 \pmf 0.3$ &        &       \\
           &      &            &      &          &                   &  0.86 &  56850 & $0.53 \pm 0.04$ & $19.2 \pmf 1.5$ &        &       \\
           &      & HD 128917  & 5.16 & F5V      & $0.378 \pm 0.075$ &  0.89 &  56849 & $0.90 \pm 0.12$ & $ 5.5 \pmf 0.6$ &        &       \\
           &      & HD 133869  & 3.72 & K3III    & $1.043 \pm 0.015$ &  0.78 &  56849 & $0.81 \pm 0.05$ & $ 6.8 \pmf 0.5$ &        &       \\
           &      &            &      &          &                   &  0.81 &  56850 & ---             & $ 3.7 \pmf 0.0$ &        &       \\
\hline
GJ 581     & 6.09 & HD 136713  & 5.83 & K2V      & $0.332 \pm 0.066$ &  0.75 &  56849 & $0.91 \pm 0.17$ & $ 5.2 \pmf 1.4$ &        &       \\
           &      &            &      &          &                   &  0.81 &  56850 & $0.48 \pm 0.03$ & $18.4 \pmf 1.1$ &        &       \\
\hline
\end{tabular}
\end{table*}
\begin{table*}
\contcaption{Observing log}
\def\pmf{\pm\phantom{0}}
\begin{tabular}{lc lclcc c crcc}
\hline
\multicolumn{2}{c}{science target} & \multicolumn{5}{c}{calibrator} & night & \multicolumn{4}{c}{conditions} \\
\multicolumn{1}{c}{name} & $H$ & \multicolumn{1}{c}{name} & $H$ & \multicolumn{1}{c}{sp. type} & $\vartheta_\mathrm{cal}$ & weight & night & seeing & \multicolumn{1}{c}{$\tau_0$} & \multicolumn{2}{c}{ins. limit} \\
 & [mag] &  & [mag] &  & [mas] & [MJD] &  & [arcsec] & \multicolumn{1}{c}{[ms]} & sci. & cal. \\
\hline
GJ 628     & 5.37 & HD 148427  & 4.88 & K0III/IV & $0.546 \pm 0.108$ &  0.90 &  56890 & $1.24 \pm 0.11$ & $11.8 \pmf 1.4$ &        &       \\
           &      & HD 148967  & 6.16 & F2/3V    & $0.256 \pm 0.051$ &  0.90 &  56890 & $1.32 \pm 0.14$ & $11.1 \pmf 1.4$ &        & $>$ lim\\
           &      & HD 149287  & 5.82 & K0V      & $0.319 \pm 0.063$ &  0.97 &  56889 & $0.81 \pm 0.06$ & $15.2 \pmf 1.2$ &        &       \\
\hline
GJ 667C    & 6.32 & HD 148729  & 6.22 & G0V      & $0.246 \pm 0.049$ &  0.80 &  56889 & $0.74 \pm 0.06$ & $34.3 \pmf 3.9$ &        &       \\
           &      & HD 155259  & 5.57 & A0/1V    & $0.242 \pm 0.048$ &  0.74 &  56849 & $0.86 \pm 0.10$ & $ 5.3 \pmf 0.8$ &        &       \\
           &      &            &      &          &                   &  0.76 &  56850 & ---             & $19.9 \pmf 1.3$ &        &       \\
           &      & HD 157338  & 5.58 & F9.5V    & $0.331 \pm 0.066$ &  0.95 &  56849 & $0.95 \pm 0.08$ & $ 4.3 \pmf 0.4$ &        &       \\
           &      &            &      &          &                   &  0.96 &  56850 & ---             & $19.2 \pmf 0.7$ &        &       \\
           &      & HD 175501  & 6.54 & F3IV/V   & $0.192 \pm 0.038$ &  0.34 &  56889 & $0.92 \pm 0.04$ & $21.5 \pmf 2.1$ &        & $\approx$ lim\\
\hline
GJ 674     & 5.15 & HD 157555  & 5.91 & F6/7V    & $0.270 \pm 0.054$ &  0.94 &  56890 & $1.30 \pm 0.11$ & $12.2 \pmf 1.3$ &        & $\approx$ lim\\
           &      & HD 159285  & 5.44 & K1III    & $0.430 \pm 0.085$ &  0.97 &  56890 & $1.54 \pm 0.26$ & $11.0 \pmf 1.5$ &        & $\approx$ lim\\
\hline
GJ 729     & 5.66 & HD 148729  & 6.22 & G0V      & $0.246 \pm 0.049$ &  0.27 &  56889 & $0.74 \pm 0.06$ & $34.3 \pmf 3.9$ &        &       \\
           &      & HD 174309  & 5.34 & F2IV     & $0.350 \pm 0.070$ &  0.92 &  56890 & $1.11 \pm 0.13$ & $13.7 \pmf 2.3$ &        &       \\
           &      & HD 175501  & 6.54 & F3IV/V   & $0.192 \pm 0.038$ &  0.97 &  56889 & $0.84 \pm 0.09$ & $24.5 \pmf 3.5$ &        &       \\
           &      & HD 179024  & 6.23 & G5III    & $0.268 \pm 0.053$ &  0.97 &  56889 & $0.74 \pm 0.08$ & $28.5 \pmf 3.1$ &        &       \\
\hline
GJ 785     & 3.58 & HD 196387  & 3.52 & K4III    & $1.085 \pm 0.015$ &  0.56 &  56210 & $0.80 \pm 0.07$ & $ 6.2 \pmf 0.7$ &        &       \\
\hline
GJ 832     & 4.77 & HD 202704  & 5.75 & G8III/IV & $0.349 \pm 0.069$ &  0.96 &  56596 & $0.64 \pm 0.05$ & $ 8.3 \pmf 0.7$ &        &       \\
           &      & HD 205048  & 4.66 & K1III    & $0.636 \pm 0.126$ &  0.86 &  56210 & $0.76 \pm 0.09$ & $ 7.0 \pmf 1.2$ &        &       \\
           &      & HD 207400  & 4.32 & K0III    & $0.651 \pm 0.129$ &  0.86 &  56210 & $0.77 \pm 0.06$ & $ 6.6 \pmf 0.5$ &        &       \\
           &      & HD 219531  & 4.29 & K0III    & $0.694 \pm 0.138$ &  0.39 &  56210 & $0.57 \pm 0.03$ & $ 8.7 \pmf 0.4$ &        &       \\
           &      & HD 221507  & 4.67 & B9.5III  & $0.354 \pm 0.070$ &  0.61 &  56210 & $0.56 \pm 0.02$ & $ 8.5 \pmf 0.2$ &        &       \\
\hline
GJ 876     & 5.35 & HD 190     & 5.38 & G8III    & $0.431 \pm 0.085$ &  0.60 &  56889 & $0.75 \pm 0.13$ & $22.4 \pmf 3.1$ &        &       \\
           &      & HD 212587  & 5.85 & G8/K0III & $0.324 \pm 0.064$ &  0.72 &  56542 & $0.87 \pm 0.05$ & $ 7.7 \pmf 0.7$ &        &       \\
           &      & HD 215097  & 4.93 & K0III    & $0.514 \pm 0.102$ &  0.84 &  56210 & $0.64 \pm 0.06$ & $ 7.5 \pmf 0.8$ &        &       \\
           &      & HD 215709  & 5.61 & F6V      & $0.313 \pm 0.062$ &  0.76 &  56542 & $1.05 \pm 0.12$ & $ 6.2 \pmf 0.7$ &        &       \\
           &      &            &      &          &                   &  0.93 &  56889 & $0.65 \pm 0.10$ & $25.2 \pmf 3.9$ &        &       \\
           &      & HD 215874  & 5.54 & F0V      & $0.300 \pm 0.060$ &  0.98 &  56253 & $0.61 \pm 0.04$ & $12.4 \pmf 0.9$ &        &       \\
           &      & HD 216357  & 5.90 & F6V      & $0.281 \pm 0.056$ &  0.61 &  56542 & ---             & $ 5.3 \pmf 0.0$ &        &       \\
           &      & HD 216402  & 5.62 & F8V      & $0.338 \pm 0.067$ &  0.86 &  56890 & $0.83 \pm 0.12$ & $15.8 \pmf 2.3$ &        &       \\
           &      & HD 217681  & 5.64 & K0III    & $0.362 \pm 0.072$ &  0.97 &  56889 & $0.58 \pm 0.05$ & $28.7 \pmf 2.2$ &        &       \\
           &      &            &      &          &                   &  0.87 &  56890 & $0.91 \pm 0.13$ & $13.5 \pmf 2.4$ &        &       \\
           &      & HD 218071  & 4.94 & K0III    & $0.570 \pm 0.113$ &  0.81 &  56210 & $0.60 \pm 0.04$ & $ 7.8 \pmf 0.7$ &        &       \\
           &      & HD 224949  & 4.92 & K0III    & $0.556 \pm 0.110$ &  0.29 &  56889 & $0.72 \pm 0.05$ & $23.6 \pmf 1.6$ &        &       \\
\hline
GJ 887     & 3.61 & HD 190     & 5.38 & G8III    & $0.431 \pm 0.085$ &  0.25 &  56850 & ---             & $17.8 \pmf 0.0$ &        &       \\
           &      & HD 214623  & 4.05 & K2/3III  & $0.776 \pm 0.011$ &  0.78 &  56597 & $1.21 \pm 0.27$ & $ 4.4 \pmf 1.1$ &        &       \\
           &      &            &      &          &                   &  0.63 &  56849 & $1.17 \pm 0.28$ & $ 3.7 \pmf 0.9$ &        &       \\
           &      &            &      &          &                   &  0.80 &  56850 & ---             & $18.1 \pmf 0.3$ &        &       \\
           &      &            &      &          &                   &  0.98 &  56889 & $0.69 \pm 0.07$ & $25.9 \pmf 3.1$ &        &       \\
           &      & HD 215616  & 4.78 & G8/K0III & $0.566 \pm 0.112$ &  0.97 &  56889 & $0.75 \pm 0.04$ & $22.8 \pmf 0.9$ &        &       \\
           &      &            &      &          &                   &  0.88 &  56890 & $0.76 \pm 0.11$ & $17.1 \pmf 2.6$ &        &       \\
           &      & HD 215678  & 5.00 & K0III    & $0.549 \pm 0.109$ &  0.85 &  56597 & $1.15 \pm 0.28$ & $ 4.4 \pmf 0.9$ &        &       \\
           &      &            &      &          &                   &  0.74 &  56849 & $1.33 \pm 0.30$ & $ 3.2 \pmf 0.7$ &        &       \\
           &      &            &      &          &                   &  0.89 &  56850 & ---             & $17.9 \pmf 0.2$ &        &       \\
           &      &            &      &          &                   &  0.90 &  56890 & $0.85 \pm 0.12$ & $15.6 \pmf 2.2$ &        &       \\
           &      & HD 216988  & 5.17 & K0III    & $0.468 \pm 0.093$ &  0.91 &  56597 & $1.00 \pm 0.11$ & $ 5.0 \pmf 0.5$ &        &       \\
           &      & HD 221750  & 4.81 & K1III    & $0.574 \pm 0.114$ &  0.98 &  56889 & $0.65 \pm 0.04$ & $26.3 \pmf 1.5$ &        &       \\
           &      & HD 224949  & 4.92 & K0III    & $0.556 \pm 0.110$ &  0.26 &  56597 & $0.91 \pm 0.05$ & $ 5.6 \pmf 0.3$ &        &       \\
           &      &            &      &          &                   &  0.35 &  56850 & ---             & $17.9 \pmf 0.1$ &        &       \\
\hline
GJ 1061    & 7.01 & HD 22000   & 7.13 & G8III    & $0.187 \pm 0.037$ &  0.93 &  56889 & $0.99 \pm 0.20$ & $20.7 \pmf 3.7$ & $>$ lim & $>$ lim\\
           &      &            &      &          &                   &  0.88 &  56890 & $0.70 \pm 0.13$ & $14.5 \pmf 2.2$ &        & $\approx$ lim\\
           &      & HD 22865   & 6.78 & F7V      & $0.188 \pm 0.037$ &  0.90 &  56890 & $0.88 \pm 0.07$ & $12.8 \pmf 1.1$ & $>$ lim & $\approx$ lim\\
           &      & HD 6022    & 7.15 & F2V      & $0.154 \pm 0.031$ &  0.52 &  56889 & $0.89 \pm 0.02$ & $19.9 \pmf 1.0$ & $>$ lim & $>$ lim\\
\hline
\end{tabular}
\end{table*}

\begin{table}
\caption{Tuning of the parameters in the interpolation of the transfer function for all setups (Sect.~\ref{sec:tf}, Eqs.~\ref{eq:tf},~\ref{eq:tfw}). MJD: Modified Julian Day; \#: Number of spectral setups; DIT: total integration time during a scan; \Nfowl: number of Fowler (i.e. non-destructive) reads of the infrared detector per scan position; \Nopd:  number of scan positions; \timescale: the time-scale to determine the weight of calibrators taken close in time; \angscale: the alt+az angular distance scale to determine the weight of calibrators close by in the sky; \minerr: the minimum relative error in the calibrator visibility considered when determining the weight of a calibrator. The first setup is a good example of an unstable night when only calibrators very close in time to the science observation ($\lesssim 20$\,min and 15\,deg) have a significant weight.  Many nights are stable enough and free of alt+az polarisation effects (e.g. MJD 56384--56850).}
\label{tab:setups}
\centering
\begin{tabular}{lrccccc}
\hline
MJD       & \# & DIT  & \Nfowl$\times$\Nopd & \timescale & \angscale & \minerr\\
$56000 +$ &    & [s]  &                     & [h]        & [deg]     & [\%]\\
\hline
210       &  1 & 0.39 &      4$\times$512   & 0.3        & 15        & 2\\
          &  1 & 0.60 &      1$\times$1024  & 0.8        & 20        & 1\\
252       &  4 & all  &            all      & 0.8        & $+\infty$ & 1\\
253       &  1 & 0.23 &      1$\times$512   & 0.8        & 20        & 1\\
325--326  &  5 & all  &            all      & 0.8        & $+\infty$ & 1\\
384--850  & 21 & all  &            all      & 0.8        & 20        & 1\\
889       &  3 & all  &            all      & 0.8        & $+\infty$ & 1\\
890       &  3 & all  &            all      & 0.8        & 20        & 1\\
\hline
\end{tabular}
\end{table}

In a companion paper by \citet{RAB18}, we have needed to obtain accurate stellar diameters of marginally resolved M~dwarfs in order to calibrate the mass-radius relation down to the fully convective regime.  While the present paper focuses on the data processing method that leads to reliable uncertainties, we find useful to show our main findings on actual data.  Our sample consists of 20 under-resolved late-type stars of the solar neighbourhood, 13 M dwarfs of the original programme by Rabus et al. plus 7 backup targets acquired during the observing campaign. 

Table~\ref{tab:obslog} summarises the science and calibrator observations that we have carried out.  The observation strategy was to observe each science target with different calibrators and on different nights.  Each science observation was bracketed by calibrator observations close in time ($\approx 10$\,min) and altitude + azimuth (alt + az) positions (a few degrees when possible).  Calibrators were chosen so that as to minimise uncertainties on the calibration of the transfer function, i.e. either unresolved or resolved with a small diameter uncertainty.  We performed the selection with the \texttt{searchCal} tool \citep{searchCal} provided by the Jean-Marie Mariotti Center (JMMC).  All calibrators have indirect (i.e. non-interferometric) diameter determinations, so they are immune to the biases investigated in this paper.  We used the Calibrator stars for 200\,m baseline interferometry \change{by} \citet[][hereafter MER05, \change{using the} method: absolute spectro-photometric calibration]{MER05} for 15 large ($\sim 1$\,mas) K0III-K5III calibrators with a typical precision of the order of 1--2\% on diameter and $\lesssim 1$\% in visibility calibration, the Catalogue of Calibrator Stars for LBSI by \citet[][\change{using the} same method]{BOR02} for one K1III star with a similar level of precision, and the JMMC Stellar Diameters Catalogue \citep[JSDC,][\change{using the} method: photometric calibration]{JSDC17} for 65 smaller ($\lesssim 0.5$\,mas) calibrators with a typical precision of 10--20\% in diameter and $\lesssim 2$\% in visibility. \citet{SWI17} have showed that the indirect spectro-photometric calibration method for large calibrators is not biased, as their catalogue is consistent both with interferometric measurements and catalogue by MER05.  For most of our science stars (15 of 20) we observed small calibrators from JSDC or ones smaller than the target from MER05.  For 4 of our science targets (GJ~1, GJ~54.1, GJ~86, GJ~370) we used one or more calibrators from MER05 that are more resolved than the target, but we also included several smaller ones from JSDC to mitigate the possible impact of a large, unexpected error in a calibrator's diameter: GJ~1 and GJ~54.1, main targets of \citet{RAB18}, have been observed together with 8 and 9 different JSDC calibrators, respectively, in addition to the 3 and 1 from MER05. While the bracketing calibrators have the largest impact on the calibration of a given science observation, other calibrators taken for other targets on the same night and with the same instrumental setup contribute to some extent to the transfer function, typically if they were taken within an hour of the science target and relatively close in the sky.  For any given target and observing night, Table~\ref{tab:obslog} lists all relevant calibrators with their relative weight (see Eq.~\ref{eq:tfw} in Sect.~\ref{sec:tf} for the weighing as a function of distance in time and position) as well as sky conditions.

Table~\ref{tab:setups} gives an overview of the instrumental setups used and the fine-tuning of calibration parameters.

\section{Data processing}
\label{sec:dataproc}

We call ``data set'', with index $\obsi$ ($\obsi$ in $1 \ldots \Nobs$), a set of $\Nint \sim 10^2$ interferograms \interf{}, indexed by \inti{} ($1 \ldots \Nint$), taken in quick succession with a single telescope pair in a single spectral channel. An interferogram is the temporal scan as function of optical path difference (OPD). Each data set of a science target will result in exactly one calibrated squared visibility amplitude \calvis, which we will refer to as ``visibility''.  The interferograms of a data set are assumed to share the projected baseline length $\base = \telsep\cdot\pos / \lambda_{\obsi}$, where $\lambda_{\obsi}$ is the effective wavelength of the spectral channel, $\telsep$ is the mean separation between the telescopes, and $\pos$ the radial unit vector representing the target's location in the horizontal coordinate system (elevation and azimuth). 

An ``observation'' consists of several data sets taken at the same time, with different values of \base, for several baselines are used at the same time, sometimes also different wavelengths. In the present case, we have used PIONIER with six telescope pairs and one to three spectral channels, so each observations consist of 6 or 18 data sets.  In a single telescope pointing, five observations are usually performed in a row (for a total of $5$--10\,min). Over the observing runs, we have acquired data for a few dozens of pointing positions per scientific target, collecting of the order of one thousand data sets per star.

A ``setup'' is a unique combination of configuration of the telescope array (stations used), instrumental setup (spectral dispersion, readout mode, scanning speed) and observing night.  Because of the span in stellar brightness among the sources and the varying observing conditions, a few setups are used each night.  In a setup, several calibrator stars and one or more scientific targets are observed. All data take\change{n} with the same setup can show some level of correlated systematics due to the wavelength calibration error.

We shall call ``\emph{baseline}'' a unique combination of a telescope pair and a setup.  In particular, we will consider that data sets taken with the same stations with different spectral configurations or on different nights originate from different baselines as they are likely to have different systematic error terms.
  
The PIONIER data reduction software \citep[hereafter \pndrs, see Sect. 5 of][]{PIONIER}, that we have modified, determines uncalibrated visibilities, computes the instrumental transfer function for calibrators of known diameters, interpolates it for scientific targets taken with the same setup, and derives the calibrated visibilities for these sources.   

Our treatment of uncertainties proceeds in four steps: we determine the ``statistical'' errors that can be inferred from the noise in the data and the uncertainty on the diameters of the calibrators (Sect.~\ref{sec:staterror}), we model an additional error term to account for the dispersion of the reduced visibilities at each baseline (Sect.~\ref{sec:bl-staterr}), introduce a highly correlated, systematic error term to account for the discrepancy between the reduced visibilities of different baselines (Sect.~\ref{sec:syserr}), and model the wavelength calibration error as an additional systematic error term (Sect.~\ref{sec:waveerr}).  Sect.~\ref{sec:covmat} gives the resulting covariance matrix.

\subsection{``Statistical'' errors}
\label{sec:staterror}

\subsubsection{Raw visibilities}
\label{sec:rawvis}

The ``raw'' visibility (uncalibrated visibility) is determined by correcting the fringe contrast of the interferograms from different atmospheric and instrumental effects such as the finite bandwidth and the flux imbalance between the beams. For the sake of clarity, we will assume that it is obtained separately on each interferogram and then averaged. However, the details of how \pndrs{} computes visibility amplitudes may vary according to setup and processing mode \citep[see Sect.~5 of][for further details of the data processing]{PIONIER}.  The bootstrap method will work as long as the resulting visibility is a function of a significant number of scans or frames. For instance, it fails for the science detector of GRAVITY with long exposures \citep{GRAVITYpipe}, but it should work for all other VLTI instruments.

Uncertainties are determined by the bootstrap method. The bootstraps $\interfboot$ ($\booti$ in $1\ldots\Nboot$) are $\Nboot$ sets of $\Nint$ interferograms picked at random with repeats from the original set $\interf$. With the exception that the first bootstrap is the original set, i.e. $\interfboot[1] = \interf$, we pick $\Nint(\Nboot - 1)$ independent random numbers $r_{\inti\booti}$ uniformly in $1\ldots\Nint$, so that $\interfboot = \interfname_{\obsi r_{\inti\booti}}$ for $\inti \ge 2$.  The random numbers $r_{\inti\booti}$ are the same for all data sets of the same observation, so that cross-channel and cross-baseline correlations are correctly measured.
 
We then obtain $\Nboot$ raw visibilities $\rawvisboot$ ($1 \le \booti \le \Nboot$) by averaging the visibility $\visibproc(\interfboot)$ obtained for each interferogram.  Together with its uncertainty, it is given by
\begin{subequations}
\begin{align}
\rawvisboot  &= \left < \visibproc(\interfboot) \right>_{\inti} \\
\drawvis     &= \frac{1}{\Nboot} \sqrt{
                   \sum_{\booti}
                   \left(\rawvisboot - \left< \rawvisboot[\bootj] \right>
                    _{\bootj} \right)^2
                }
\end{align}
\label{eq:rawvisboot}
\end{subequations}

In this work, we picked $\Nboot = 5\times10^3$ so that we can derive a covariance matrix for a few thousands of data sets and use the (very slightly biased) sample variance to avoid numerical issues. The original version of \pndrs{} published by \citet{PIONIER} also bootstraps the interferograms (with $\Nboot \sim 10^2$) to determine the uncertainty, but discards them afterwards. It keeps the value $\rawvisboot[1] \pm \drawvis$ of Eq.~(\ref{eq:rawvisboot}) and propagates the errors assuming an uncorrelated multivariate Gaussian distribution. We have modified the software to keep all the bootstraps down to the final product and get an empirical sampling of the calibrated visibility distribution.

\subsubsection{Instrumental transfer function}
\label{sec:tf}
There are still instrumental effects, difficult to compute, in the raw visibilities, so that the fringe contrast is lower than expected from a theoretical point of view. To remove them, the transfer function (also known as instrumental visibility) is calculated on unresolved sources or targets of precisely known geometry, observed in the vicinity of the scientific targets.  Then it is interpolated for science targets. Some care has to be taken to include the uncertainties on the calibrators' diameters and variations of the transfer function due to changing atmospheric conditions.

\def\obsi{c}
If the star is a calibrator with known uniform disc diameter $\ud \pm \dud$, then $\Nboot$ diameters $\udboot$ are picked at random assuming a Gaussian distribution $\Norm(\ud, \dud^2)$, with the exception that $\udboot[1] = \ud$.  This is done once per calibrator star for all the observing runs, so that correlations from calibrator errors are correctly propagated to the final data. 

For data set $\obsi$ corresponding to a calibrator and bootstrap number $\booti$, the ratio of the raw visibility $\rawvisboot$ to the theoretical uniform disc visibility $\udvisboot$ yields the transfer function $\tfboot$:
\begin{subequations}
\begin{align}
\udvisboot  &= \udvisfunc\left(\pi\base\udboot\right)\\
\dudvis     &= \deriv{\udvisfunc}{x} (\pi\base\ud) \pi\base\dud\\
\tfboot     &= \frac{\rawvisboot}{\udvisboot }\\
\dtf        &= \tfboot[1] \sqrt{
                    \left(\frac{\drawvis}{\rawvis}\right)^2
                  + \left(\frac{\dudvis}{\udvisboot[1]}\right)^2
               }  
\end{align}
\end{subequations}
where the theoretical visibility amplitude of a uniform disc is given by the function $\udvisfunc(x) = |2 J_1(x) / x|^2$ with $J_1$ the Bessel function of the first kind.
\def\obsi{d}

If the star is a scientific target, the reduction software interpolates the transfer function using calibrator observations $c_1, \ldots, c_n$ obtained on the baseline (same telescope pair and setup).  Calibrator observations close in time and/or position in the sky and with smaller error bars are given more weight. If, for bootstrap number $\booti$, the transfer functions $\tfname_{c_1\booti}, \ldots, \tfname_{c_n\booti}$ are determined at time $t_{c_1}, \ldots, t_{c_n}$ and horizontal coordinates $\posname_{c_1}, \ldots, \posname_{c_n}$, then the estimated transfer function for a science observation $\obsi$ at time $t_\obsi$ and position $\pos$ is given by
\begin{subequations}
\begin{align}
        \tfboot &= \frac{\sum_k w_k \tfname_{c_k\booti}}{\sum_k w_k}
          \label{eq:tf}\\
        w_k &= 
            \max\left(\minerr,  \frac{\Delta\tfname_{c_k\booti}}
                                     {      \tfname_{c_k\booti}} \right) ^ {-2}
            \mathrm{e} ^ {-\frac{(t_{c_k} - t_{\obsi})^2 }{\timescale^2}} 
            \mathrm{e} ^ {-\frac{\angdist (\posname_{c_k}, \pos)^2}{\angscale^2}} \label{eq:tfw}
\end{align}
\end{subequations}
where $\alpha(\posname_1, \posname_2)$ is the alt + az difference between telescope pointing positions $\posname_1$ and $\posname_2$. $\timescale$, $\angscale$, and $\minerr$ are constants within a given setup. $\timescale$ is the time-scale of variations of the transfer function, typically of the order of 1~hour in the original \pndrs{} software, but we have shortened it for a few very ``agitated'' nights (see Table~\ref{tab:setups}). $\angscale$ is the angular size on the sky over which the transfer function varies.  In the original software, $\angscale = +\infty$ so that calibrators have a weight independent of their position relative to the science target. However, in some nights, polarisation effects led us to use a finite value (see Table~\ref{tab:setups}). $\minerr = 0.01$ (also used in the original \pndrs) is the minimum relative uncertainty we consider in the determination of the weight of calibrator observations, in order to avoid that a calibrator with an unexpectedly low uncertainty biases the transfer function. 

The calculation of $\dtf$ includes three terms\change{:}
\begin{enumerate}
	\item the error propagation using Eq.~ (\ref{eq:tf}), which includes the uncertainty on the diameter of the calibrators;
    \item an interpolation uncertainty taking into account the varying atmospheric conditions between target and calibrator, which we measure by comparing the interpolated transfer function for calibrator observations with the measured one;
    \item an extrapolation uncertainty for data not bracketed by calibrators (which we have avoided).
\end{enumerate}
An example for the first two uncertainty terms of the transfer function is given in Sect.~3 of \citet{NUN17} in the case of weights linear in time.  The full expression for $\dtf$ is not given here for it follows exactly the same steps as in the original software.

\subsubsection{Calibrated visibilities}
\label{sec:calvis}

The calibrated visibilities \calvisboot, their uncertainties \dcalvis, and their covariances \covar{} are given by:
\begin{subequations}
\begin{align}
    \calvisboot{} &= \frac{\rawvisboot}{\tfboot} \label{eq:calvisboot},\\
    \dcalvis      &= \calvisboot[1]{} \sqrt{
                    \left(\frac{\drawvis}{\rawvis}\right)^2
                  + \left(\frac{\dtf}{\tfboot[1]}\right)^2
                    } \label{eq:dV}\\
    \covar &= \frac{
        \sum_{\booti} 
        \left(\calvisiboot - \left<\calvisiboot[\bootj]\right>_\bootj \right)
        \left(\calvisjboot - \left<\calvisjboot[\bootj]\right>_\bootj \right)
         }{\Nboot}\label{eq:covar}.\\
\intertext{In some cases, the uncertainties on the calibrated visibility obtained from the standard deviation of the calibrated bootstraps}
    \dcalvisboot &= \sqrt{\covaridiag} \label{eq:var}
\end{align}
\end{subequations}
differ to some extent from the ones derived by error propagation ($\dcalvis$ in Eq.~\ref{eq:dV}) in the original software.  The reasons are correlations, departure from the Gaussian distribution, and non-linearity in the calculations in the reduction software (in particular the divisions).  

\subsection{Baseline-dependent ``statistical'' errors}
\label{sec:bl-staterr}

We have noted that within a given baseline, it frequently happens that the dispersion of the data points is higher than what our carefully deduced uncertainties suggest. Because most of the stars in our programme are \textit{bona fide} centro-symmetric targets, under-resolved in the $H$ band at VLTI, the uniform disc model \citep[or any symmetric model, see][]{LAC03} should fit the data correctly.  Thus, we expect the reduced chi squared of a least squares fit to the data of a single baseline to be close to unity.  While it is the case on some baselines, it might be significantly higher ($\approx 2$) on others. 

It is quite possible that the determination of the instrumental transfer function in quickly changing conditions is not perfect. In particular, the interpolation of the transfer function (Eq.~\ref{eq:tf}) would fail if conditions changed abruptly.  In general, our knowledge on the variation of the transfer function is very limited so we rely on an imperfect, generic smoothing law (Eq.~\ref{eq:tfw}).

We also considered that time-correlations may impact the bootstrap method.  In Sect.~\ref{sec:rawvis}, the determination assumes that the $\visibproc(\interf)$, for $\inti$ in $1 \cdots \Nint$, are not time-correlated. Unfortunately, the data are too noisy to infer a meaningful auto-correlation function and model its impact on the result. However, we know that the observing cadence ($\sim 1$\,s) is significantly slower than the typical interferometric coherence time \citep[$\sim 100$\,ms in the IR, see][]{PER97,DI03,GLI11} and atmospheric turbulence time $\tau_0$ ($\sim 10$\,ms in the IR). So, we expect little correlation from the differential atmospheric piston and fibre injection variations.  In the case unsuspected sources of temporal correlation did show up, we would expect to underestimate uncertainties. However, we have checked that, even with highly correlated visibilities, the impact is very small: in the case of a correlation coefficient of 0.9 between consecutive interferograms, we simulated batches of 100 measurements, which yielded a bootstrap estimate of $\drawvis$ within 10\% of the correct value. 

In the absence of a clear understanding of these errors, we decide to model them as an uncorrelated additive term $\sysuncorr$.  It has expectancy 0, standard deviation $\sigmauncorr\calvis$, and correlations 0.  For each baseline where the reduced chi squared $\chisqr$ differs significantly from 1\footnote{$\chisqr$ is expected to be $1.00 \pm 0.22$ with a 3-$\sigma$ confidence interval for an observation with two telescope pointings and three spectral channels, but we have found values of 2 to 4 on some baselines.}, the value of $\sigmauncorr$ is adjusted so that a least-squares fit to the data of this baseline (and only these data) yields a reduced $\chisqr = 1$.

\begin{table*}
    \caption{The different uncertainty models (Sect.~\ref{sec:ud}).  The first four models used uncorrelated statistical errors, the three following ones correlated  statistical errors, and the last ones include correlated systematic errors. 
    \emph{Data uncertainties} can be obtained from \pndrs{} error propagation, the bootstrap variances, the bootstrap variances plus a noisy correlation matrix representing uncorrelated data (``corr. noise''), or the bootstrap covariances.
    \emph{Baseline-dependent statistical errors} of uncertain origin may be ``ignored'' or ``fit'' on each baseline so that baseline $\chisqr = 1$. 
    \emph{Systematic errors} can either be modelled with ``correlation'' matrix or one can just perform a ``rescaling'' of all statistical errors to a global $\chi^2 = 1$. For the sake of comparison we also include the incorrect correlation matrix (``PPP corr.'') that may lead to Peelle's Pertinent Puzzle.
    \emph{Fitted visibilities} can either be the mean of the 
bootstraps (``mean'') or all bootstraps (``bootstraps'').  In the bootstrap case, the diameter value and uncertainty are obtained from the median and 1-$\sigma$ confidence interval of the obtained distribution.
    \emph{Mathematical expression for uncertainties}: either an error vector or a covariance matrix.
       }
    \label{tab:err-model}
    \begin{tabular}{l lll lll}
    \hline\hline
    Model          & Data uncertainties  & Baseline-dependent  & Systematic  & Fitted       & \multicolumn{2}{l}{Mathematical expression}\\
                   &                     & statistical errors  & errors      & visibilities & \multicolumn{2}{l}{used for the uncertainties}\\
    \hline                                                                  
    \modelpndrs    & propagation         & ignored             & rescaling   & mean      & Eq.~(\ref{eq:dV})            & $\dcalvis$                           \\
    \modelvar      & variances           & ignored             & rescaling   & mean      & Eq.~(\ref{eq:var})           & $\dcalvisboot$                       \\
    \modelvarboot  & variances           & ignored             & rescaling   & bootstraps   & Eq.~(\ref{eq:var})           & $\dcalvisboot$                       \\
    \modelvarnoisy & variances           & ignored             & rescaling   & mean      & Eq.~(\ref{eq:noisycovar})    & $\noisycovar$                        \\
                   & + corr. noise \\ 
    \hline                                                                                                     
    \modelcov      & covariances         & ignored             & rescaling   & mean      & Eq.~(\ref{eq:covar})         & $\covar$                             \\
    \modelcovbl    & covariances         & fit                 & rescaling   & mean      & Eq.~(\ref{eq:covar})         & $\ubcovar$ with $\sigmasys = \sigmawave =  0$      \\ 
    \modelcovboot  & covariances         & fit                 & rescaling   & bootstraps   & Eq.~(\ref{eq:covar})         & $\ubcovar$ with $\sigmasys = \sigmawave =  0$      \\
    \hline                                                                                                     
    \modelppp      & covariances         & fit                 & PPP corr.   & mean      & Eq.~(\ref{eq:covar})         & $\ubcovar$ with $\meancalvis=\calvis$\\
    \modelsys      & covariances         & fit                 & correlation & mean      & Eq.~(\ref{eq:covar})         & \ubcovar                             \\
    \modelsysboot  & covariances         & fit                 & correlation & bootstraps   & Eq.~(\ref{eq:covar})         & \ubcovar                             \\
    \hline
    \end{tabular}
\end{table*}
\begin{table*}
    \caption{Uniform disc diameters of the sample using different model fitting routines of Sect.~\ref{sec:ud}: $\diampndrs$ (uncorrelated least squares using the error propagation of the reduction pipeline), $\diamvar$ (uncorrelated least squares using observed variances), $\diamcovar$ (least squares using observed covariances),  $\diamcovbl$ (least squares using observed covariance and fitting baseline-dependent statistical errors), and $\diamsys$ (correlated least squares with baseline systematics).  For each model, the following relative errors on the visibilities are given: $\sigmastat$ is obtained from the data, $\sigmauncorr$ is derived by modelling an additional baseline-dependent statistical (i.e. uncorrelated) error term, $\sigmasys$ is obtained by fitting the strength of the systematics (fully correlated on the same baseline). The reduced $\chisqr$ of the fits is also given. For GJ~581, the reduced chi squared of the \modelcovbl{} model is below one, so systematics adjust to zero and $\sigmasys = 0$ for \modelsys.}

    \label{tab:ud-comp}
    \tabcolsep=1.5pt
    \small
    \begin{tabular}{p{4em} crp{3em} crp{3em} crp{3em} crrp{3em} crrrp{3em}}
\hline
Star          & \multicolumn{3}{c}{VAR/PROP}                  & \multicolumn{3}{c}{VAR}                       & \multicolumn{3}{c}{COV}                       & \multicolumn{4}{c}{COV/BL}                                    & \multicolumn{5}{c}{SYS}                                                      \\
              & $\vartheta$   & \chisqr       & \sigmastat    & $\vartheta$   & \chisqr       & \sigmastat    & $\vartheta$   & \chisqr       & \sigmastat    & $\vartheta$   & \chisqr       & \sigmastat    & \sigmauncorr  & $\vartheta$   & \chisqr      & \sigmastat    & \sigmauncorr  & \sigmasys     \\
              & [mas]         &               & [$\%V^2$]     & [mas]         &               & [$\%V^2$]     & mas           &               & [$\%V^2$]     & [mas]         &               & \multicolumn{2}{c}{[$\%V^2$]} & [mas]         &              & \multicolumn{3}{c}{[$\%V^2$]}                 \\
\hline\hline
GL1           & 0.780(01)     &          2.52 &           3.1 & 0.782(01)     &          3.43 &           2.4 & 0.797(03)     &          3.44 &           2.4 & 0.796(02)     &          1.08 &           2.4 &           2.0 & 0.794(05)     &          1.00&           2.4 &           2.0 &           1.0 \\
GL54.1        & 0.400(07)     &          3.33 &           6.6 & 0.407(06)     &          3.31 &           6.7 & 0.408(10)     &          2.95 &           6.7 & 0.402(09)     &          1.63 &           6.7 &           4.3 & 0.397(35)     &          1.00&           6.7 &           4.3 &           4.3 \\
GL86          & 0.662(02)     &          2.12 &           4.4 & 0.648(02)     &          3.10 &           4.1 & 0.683(05)     &          3.09 &           4.1 & 0.678(04)     &          1.27 &           4.1 &           3.8 & 0.683(15)     &          1.00&           4.1 &           3.8 &           2.6 \\
GL229         & 0.865(03)     &          5.96 &           4.9 & 0.860(03)     &          6.95 &           3.5 & 0.888(05)     &          3.26 &           3.5 & 0.870(04)     &          1.27 &           3.5 &           3.3 & 0.840(12)     &          1.00&           3.5 &           3.3 &           1.6 \\
GL273         & 0.743(02)     &          1.92 &           4.9 & 0.744(02)     &          2.87 &           4.3 & 0.760(05)     &          2.80 &           4.3 & 0.752(04)     &          1.14 &           4.3 &           2.2 & 0.763(10)     &          1.00&           4.3 &           2.2 &           1.6 \\
GL370         & 0.522(02)     &          2.04 &           2.2 & 0.518(03)     &          2.73 &           2.0 & 0.536(05)     &          2.77 &           2.0 & 0.532(04)     &          1.15 &           2.0 &           1.4 & 0.530(12)     &          1.00&           2.0 &           1.4 &           1.6 \\
GL406         & 0.521(04)     &          2.02 &           6.5 & 0.522(04)     &          2.03 &           6.5 & 0.539(08)     &          2.18 &           6.5 & 0.540(07)     &          1.17 &           6.5 &           2.8 & 0.562(20)     &          1.00&           6.5 &           2.8 &           2.7 \\
GJ433         & 0.390(10)     &          2.95 &           4.5 & 0.363(11)     &          3.93 &           3.7 & 0.426(15)     &          2.91 &           3.7 & 0.422(13)     &          1.28 &           3.7 &           2.5 & 0.457(30)     &          1.00&           3.7 &           2.5 &           2.6 \\
GL447         & 0.513(07)     &          1.63 &           8.1 & 0.519(08)     &          2.73 &           4.8 & 0.539(16)     &          2.97 &           4.8 & 0.533(12)     &          1.29 &           4.8 &           6.1 & 0.524(29)     &          1.00&           4.8 &           6.1 &           2.3 \\
GL551         & 1.059(03)     &          4.04 &           3.2 & 1.051(03)     &          4.62 &           3.1 & 1.038(05)     &          3.29 &           3.1 & 1.068(04)     &          1.17 &           3.1 &           3.0 & 1.066(07)     &          1.00&           3.1 &           3.0 &           1.1 \\
GJ581         & 0.467(04)     &          1.43 &           3.7 & 0.471(05)     &          2.61 &           2.7 & 0.460(08)     &          2.15 &           2.7 & 0.464(07)     &          0.97 &           2.7 &           2.3 & 0.464(07)     &          0.97&           2.7 &           2.3 &           0.0 \\
GJ628         & 0.634(02)     &          1.96 &           1.8 & 0.636(02)     &          1.85 &           1.8 & 0.645(07)     &          2.23 &           1.8 & 0.646(05)     &          1.13 &           1.8 &           0.9 & 0.644(14)     &          1.00&           1.8 &           0.9 &           1.6 \\
GJ667C        & 0.390(08)     &          1.90 &           4.8 & 0.371(08)     &          3.01 &           3.6 & 0.401(15)     &          2.86 &           3.6 & 0.411(11)     &          1.05 &           3.6 &           3.1 & 0.406(14)     &          1.00&           3.6 &           3.1 &           0.7 \\
GJ674         & 0.705(06)     &          1.10 &           3.7 & 0.704(06)     &          1.21 &           3.7 & 0.695(18)     &          1.70 &           3.7 & 0.700(15)     &          1.07 &           3.7 &           1.4 & 0.720(37)     &          1.00&           3.7 &           1.4 &           2.6 \\
GJ729         & 0.601(03)     &          1.89 &           3.6 & 0.602(03)     &          1.81 &           3.7 & 0.620(08)     &          2.42 &           3.7 & 0.615(06)     &          1.27 &           3.7 &           0.9 & 0.625(20)     &          1.00&           3.7 &           0.9 &           2.4 \\
GL785         & 0.681(09)     &          6.64 &           2.1 & 0.630(10)     &          21.5 &           1.7 & 0.619(11)     &          7.04 &           1.7 & 0.720(08)     &          1.82 &           1.7 &           2.4 & 0.677(21)     &          1.00&           1.7 &           2.4 &           2.0 \\
GJ832         & 0.807(04)     &          2.36 &           2.5 & 0.801(04)     &          3.12 &           2.2 & 0.803(05)     &          2.21 &           2.2 & 0.798(05)     &          1.29 &           2.2 &           1.2 & 0.794(\change{10}) &          1.00&           2.2 &           1.2 &           1.0 \\
GJ876         & 0.680(02)     &          2.14 &           4.5 & 0.683(02)     &          2.49 &           4.2 & 0.692(04)     &          2.69 &           4.2 & 0.691(03)     &          1.12 &           4.2 &           3.2 & 0.686(09)     &          1.00&           4.2 &           3.2 &           1.7 \\
GL887         & 1.298(01)     &          4.01 &           2.9 & 1.297(01)     &          5.08 &           2.5 & 1.302(02)     &          4.27 &           2.5 & 1.300(02)     &          1.16 &           2.5 &           3.7 & 1.29\change{7}(04) &          1.00&           2.5 &           3.7 &           1.\change{4} \\
GL1061        & 0.292(14)     &          1.93 &           3.2 & 0.313(12)     &          2.27 &           2.9 & 0.332(18)     &          1.92 &           2.9 & 0.332(16)     &          1.16 &           2.9 &           1.9 & 0.344(34)     &          1.00&           2.9 &           1.9 &           1.8 \\
\hline
\end{tabular}

\end{table*}

\subsection{``Systematic'' errors}
\label{sec:syserr}
We have tried to avoid the commonest systematic error sources with PIONIER.  Any calibrator showing hints of binarity, thus being able to skew the transfer function for baselines parallel to the binary separation, was excluded.  Calibrators were chosen close (a few degrees when possible) to the science targets so that the differential polarisation effect is minimised.  On nights were this effect was impacting the data processing with the default \pndrs{} parameters, the instrumental transfer function was interpolated in a way that calibrators distant in alt+az position are filtered out (see Sect.~\ref{sec:tf} and Table~\ref{tab:setups}).

In spite of these efforts, there is strong evidence of systematics in our data.  Visually (see Fig.~\ref{fig:visib}, in particular GJ~86, GJ~1061, GJ~1), the data of some baselines do not align with those at other baselines.  This first look is confirmed by a reduced $\chi^2 > 1$ on most of our fits in spite of a detailed modelling of statistical errors (Sects.~\ref{sec:staterror}~\& \ref{sec:bl-staterr}).

We model these systematic errors by a multiplicative term $\sys$ with high correlations along the same baselines.  It has expectancy 1, standard deviation $\sigmasys$, and correlation coefficients $\corr \approx 1$ for all observations $\obsi$ of the same baseline, and 0 for observations on different baselines.  The value of $\sigmasys$ is adjusted so that a least squares fit to the full data set yields a reduced $\chi^2 = 1$.
 
\subsection{Wavelength calibration errors}
\label{sec:waveerr}
\citet{GAL18} showed that the PIONIER wavelength calibration has a relative uncertainty of $\approx 0.35$\% by alternating observations of a well know binary with PIONIER and wavelength-stabilised second-generation VLTI instrument GRAVITY \citep{GRAVITY}. The uncertainty on the central wavelength of spectral channels of a given spectral configuration, $\sigmawave = \Delta\lambda/\lambda$ produces an uncertainty on the projected baseline $\base$ of the same amount (0.35\%).  To model it exactly, we should introduce correlated uncertainties along the $x$-axis ($\base \pm \sigmawave\base$) in addition to those on the $y$-axis ($V^2 \pm \Delta V^2$).  Since the model is a continuously differentiable function of baseline, we decide instead to translate this small error in baseline into a small error in visibility.   For under-resolved objects, $1 - \calvis \propto  \base^2$ so that we obtain:
\begin{equation}
    \Delta V^2 \approx 2(1 - V^2) \sigmawave.
\end{equation}

We model this systematic error source as an additive term \syswave{} of zero expectancy, standard deviation $2 \sigmawave (1 - \calvis)$ and correlation $\approx 1$ for all observations taken with the same spectral setup on the same night.  Because \citet[][see their Fig.~1]{GAL18} show that the wavelength error varies from one night to the other, we have assumed that correlations are zero \change{for} data taken on different nights and/or different spectral configurations.

For observations taken on a single night in a single spectral setup, we can expect a diameter uncertainty of $0.35\%$. However, according to our observation strategy, most of the stars in our sample were observed on different nights, so we expect these systematics to have a lower impact on the final uncertainty determination (0.15 to 0.25\%).

\subsection{Final uncertainty determination}
\label{sec:covmat}

The final quantity we measure and fit is therefore
\begin{equation}
   \ubvis = \sys(\calvis + \sysuncorr) + \syswave.
    \label{eq:ubvis}
\end{equation}

The determination of the uncertainties and the covariance matrix must be done with some care. It is well known that a na\"\i{}ve error propagation in presence of high correlations can lead, and indeed leads as we discovered in our data, to a least-squares fit that falls far off the data and yields a parameter estimation inconsistent with a more careful analysis. The paradox, known as Peelle's pertinent puzzle \citep{PEE87}, has an easy remedy in the case of a fit by a constant value \citep{NEU12} or a set of constant values \citep{NEU14}. In their analytic derivation of the covariances,  the weighted average of the measurements replaces the individual measurements.  We need to generalise their results to suit our needs, since the different visibilities of the same baseline do not have the same expected value ($\udvisfunc$ is not constant).  As \citet{ZHA92} already noted, there is no longer an obvious (analytic) candidate for the weighted average in the case of a non-linear model.  We decided to use the most ``natural'' estimate: we replace the visibilities $\calvis$ by their estimator $\meancalvis$ obtained from a least-squares model fit to the data of the given baseline and setup.  In the case of a constant model, it would give the same result as \citet{NEU14}. 
\def\varname{\mathcal{V}}
\def\varsys{\ensuremath{\varname^\mathrm{sys}}}
\def\varbl{\ensuremath{\varname^\mathrm{bl}_b}}
\def\varwave{\ensuremath{\varname^\mathrm{wl}}}

We introduce the following variances related to the baseline-dependent statistical errors \sysuncorr, the systematic errors \sys{}, and the wavelength calibration errors \syswave:
\begin{subequations}
\begin{align}
   \varsys  = \Var(\sysuncorr)\phantom{\calvis} \,\, &= \big(  \sigmasys    \meancalvis       \big)^2\\
   \varbl   = \Var(\sys\calvis) &= \big(  \sigmauncorr \calvis           \big)^2\\
   \varwave = \Var(\syswave)\phantom{\calvis} \,   &= \big( 2\sigmawave   (1 - \meancalvis) \big)^2
\end{align}
\end{subequations}

The final error and correlation matrix are given by
\begin{subequations}
\begin{align}
   (\dubvis)^2       &= 
                         (\dcalvis)^2 
                       + \varsys 
                       + \varbl
                       + \varwave\label{eq:dubvis}\\
   \ubcovar       &= 
        \begin{cases}
                     \covar + \phantom{\corr}\varwave + \phantom{\corr}\varsys + \varbl 
                        \text{ if $\obsi = \obsj$}
                     \\
                     \covar + \corr\varwave + \rho\varsys\phantom{\null + \varbl }
                        \text{ if $\baselinei = \baselinej$}
                     \\
                     \covar + \corr\varwave\phantom{\null + \rho\varsys + \varbl }
                        \text{ if $\confi = \confj$}
                         \\
                     \covar\phantom{\null  + \corr\varwave  + \rho\varsys + \varbl }
                     \text{ otherwise}
        \end{cases}
        \label{eq:syscovar}
\end{align}
\end{subequations}
where $\baselinei$ and $\confi$ are the baseline and spectral setup of data set $\obsi$. In order to prevent a (numerically) singular matrix, we use $\corr = 0.95$ instead of 1.  Since errors are relatively small (a few percents) the second-order terms ($< 0.1$\%) arising when propagating the errors from Eq.~\ref{eq:ubvis} have been ignored in Eqs.~\ref{eq:dubvis}~\&~\ref{eq:syscovar}.

\section{Impact of the uncertainty model}
\label{sec:ud}

\begin{figure*}
\centering
\includegraphics{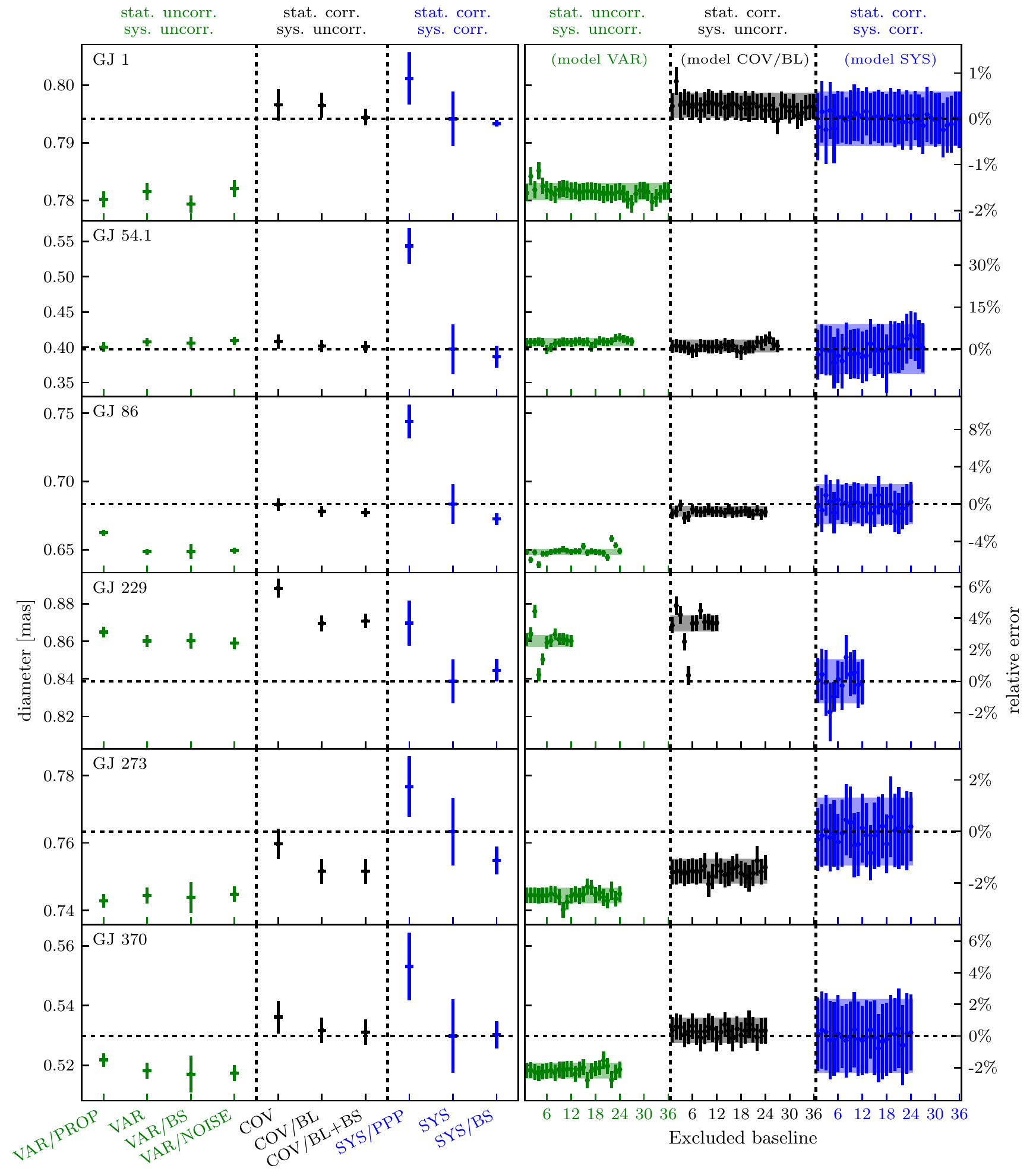}
\caption{Comparison of uniform disc diameters in milliarcseconds (mas) (left axis) and their deviation from the robust \modelsys{} model estimate in \% (right axis) obtained with different error models (Sect.~\ref{sec:ud}). For each star, two graphs are displayed on the same line.  In each graph, the least squares fits include models with uncorrelated statistical and systematic errors (green, points on the left), correlated statistical errors and uncorrelated systematics (black, points in the middle), and correlated statistical and systematic errors (blue, points on the right.)  \emph{Left graph:} influence of correlations and systematic errors. \emph{Right graph:} robustness with respect to baseline systematics, by comparing fits where a baseline has been removed.}
\label{fig:ud-comp}
\end{figure*}

\begin{figure*}
\centering
\includegraphics{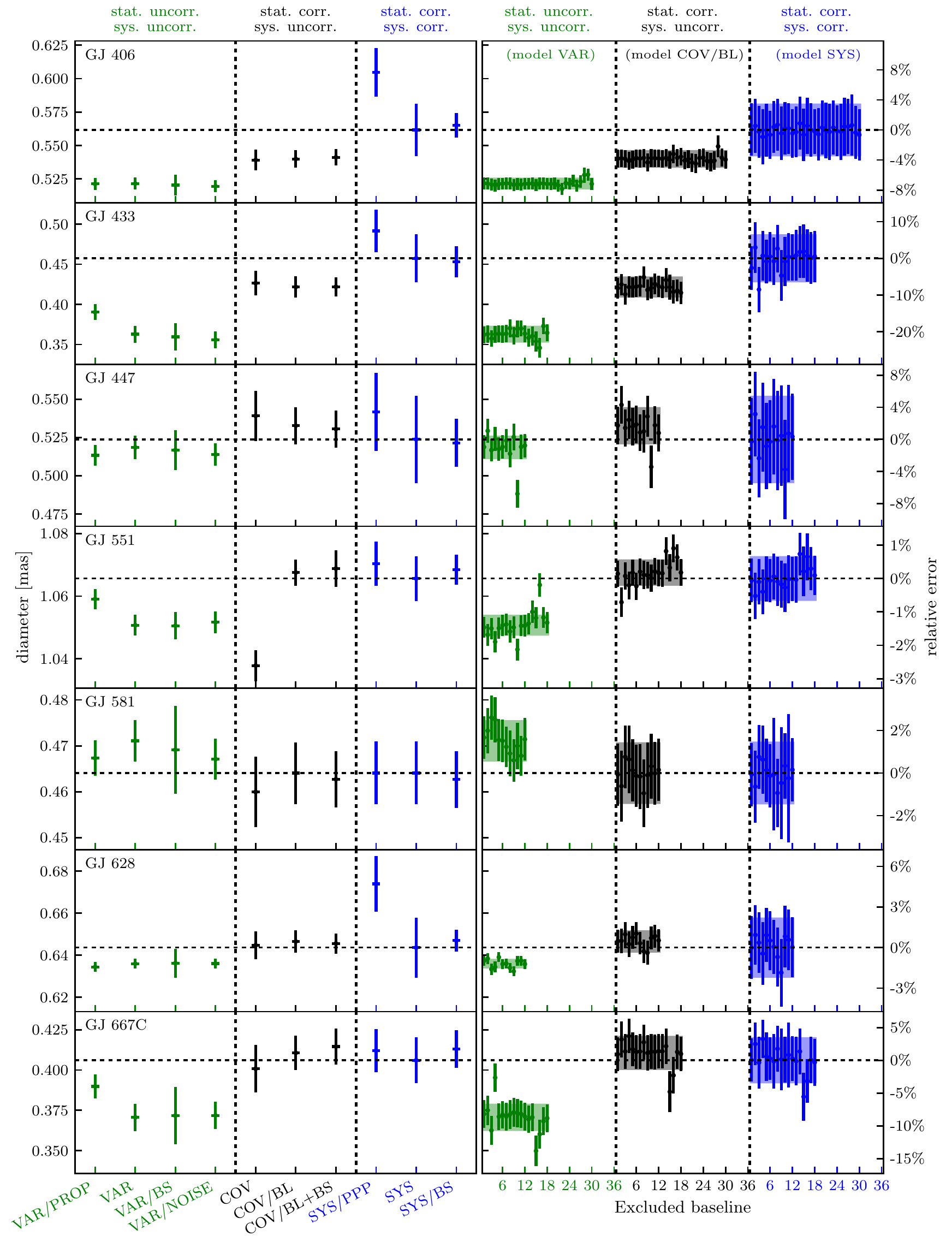}
\contcaption{Comparison of uniform disc diameters using different fitting methods.}
\end{figure*}

\begin{figure*}
\includegraphics{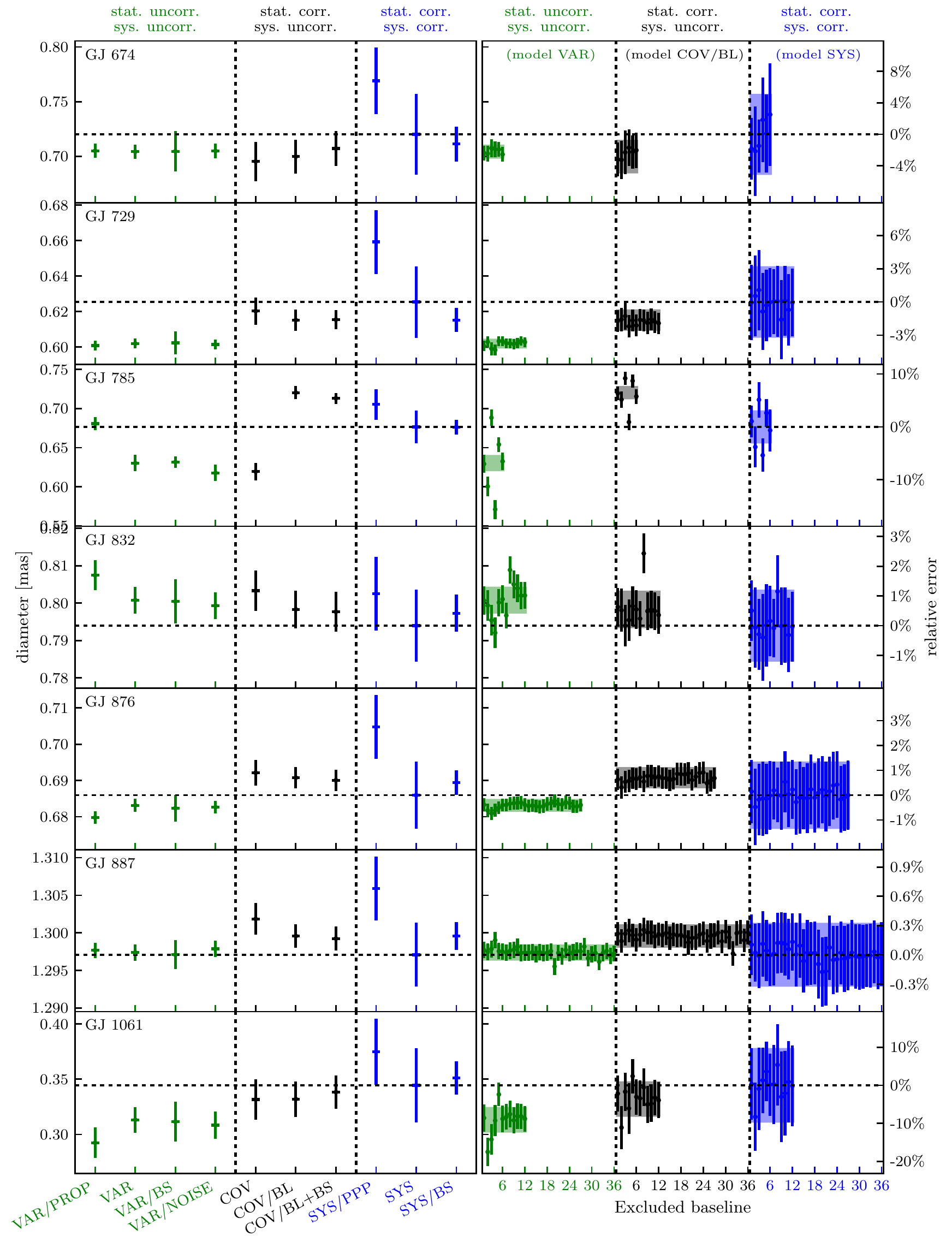}
\contcaption{Comparison of uniform disc diameters using different fitting methods.}
\end{figure*}

\begin{figure*}
\includegraphics{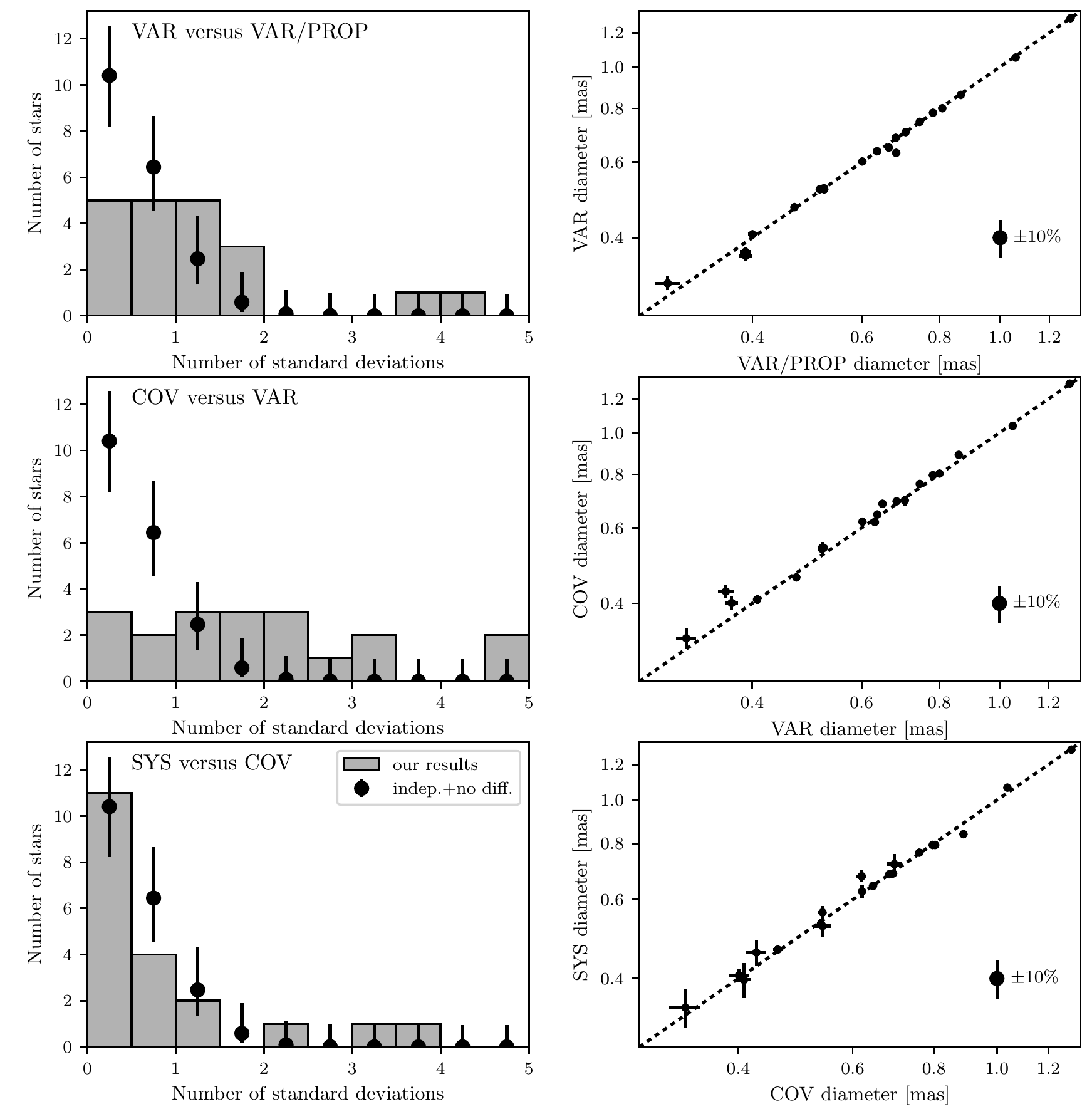}
\caption{Comparison between diameter estimates for the four error models that most differ \modelpndrs, \modelvar, \modelcov, \modelsys{} (see Sect.~\ref{sec:ud}). 
\textit{Left:} histogram of the number of standard deviations between the estimates.  As a reference, we plot the expected distribution if the two estimates had the same expected value and were independent Gaussian random variables. 
\textit{Right:} direct comparison of the diameters estimates. The dashed line indicates the place  where the estimates are equal.}
\label{fig:estim-comp}
\end{figure*}

We assess the respective importance of the correlation and the level of detail used in the determination of uncertainties by comparing the results obtained with different error models.  These error models, explained below, are briefly summarised in Table~\ref{tab:err-model}.  The values of the uniform disc diameters obtained from these models are given in Table~\ref{tab:ud-comp} and Fig.~\ref{fig:ud-comp}. For the most discrepant error models, the deviation between estimates is shown in Fig.~\ref{fig:estim-comp}.  

In all models, a least-squares fit of the value of the diameter is performed on the calibrated visibilities $\calvis$, sometimes for each bootstraps $\booti$ using $\calvisboot$. When the fits are performed on each bootstrap, the diameter estimate is the median of obtained values and the uncertainty is obtained with the 1-$\sigma$ confidence interval. 

We have gathered the different models into three groups.  In the first one uncorrelated errors (uncertainties $\dcalvis$ or $\dcalvisboot$) are assumed and no systematics are taken into account. They are called \modelvar{} (``variance''). The second group, uses correlated errors (covariance matrix $\covar$ or $\ubcovar$) but no systematics, they are called \modelcov{} (``covariance'').  The last group of models use both correlated statistical errors and systematics, they are called \modelsys{} (``systematics'').  Within these groups small differences in the error handling have been considered, they are specified after a slash.

\subsection{Error propagation} 
\label{sec:influence-prop}

The two first error models we introduce and compare take neither correlations nor systematics into account. They differ in the way the error propagation is performed from the raw to calibrated visibilities.  In the first model, \modelpndrs{}, the standard quadratic addition of error terms along the way is performed (fit to $\calvis \pm \dcalvis$).  The second one, \modelvar, determines the error from the bootstraps (fit to $\calvis \pm \dcalvisboot$). \modelpndrs{} and \modelvar{} are displayed as the first two points in the left panels of Fig.~\ref{fig:ud-comp}.

If errors are not strictly Gaussian, as expected from the quotient in Eq.~(\ref{eq:calvisboot}), the bootstrap method samples the real distribution of the calibrated visibilities and estimates correctly its average and confidence intervals, while the standard propagation may not.

In most cases (see Fig.~\ref{fig:estim-comp}, top-left panel), the estimates for the diameters using \modelpndrs{} and \modelvar{} are consistent with each other and differ by at most 2-$\sigma$.  For two stars (10\% of the sample), GJ~785\change{,} and GJ~86\change{,} the results are at least 3-$\sigma$ apart.  We conclude that departure from a Gaussian distribution has not a significant impact in most cases.

\subsection{Correlation of statistical errors}
\label{sec:influence-corr}

The \modelvar{} error model group assumes that errors on the visibilities are not correlated ($\calvis \pm \dcalvis$ is fit), while the \modelcov{} model group does ($\calvis$ is fit with correlation matrix $\covar$).  We expect that positive correlations increase the errors bars on the diameter, as the different data points become partially redundant.  They can also shift its estimated value, as groups of correlated data lose their weight relative to uncorrelated data.

This is indeed what we observe for most of our stars (in the left panel of Fig.~\ref{fig:ud-comp}, 2nd and 5th points from the left labelled \modelvar{} and \modelcov{}).  The difference is quite significant as 40\% of the stars show a discrepancy of at least 2-$\sigma$ (see Fig.~\ref{fig:estim-comp}, middle-left panel).

\subsection{Baseline-dependent statistical errors}

As explained in Sect.~\ref{sec:bl-staterr}, the data at some baselines show more dispersion than the statistical errors determined during the data reduction.  An additional baseline-dependent error term has been introduced to account for this discrepancy in model \modelcovbl.

As it can be seen in Fig.~\ref{fig:estim-comp} (for star GJ~785), including these errors can significantly modify the diameter estimate.  When a few baselines are impacted by very high noise of unknown origin, most of the time bad and fast varying atmospheric conditions, the inclusion of an additional error term decreases the weight of these baselines in the fit (for GJ~785, see two baselines around 50 and 70--80\,M$\lambda${} in Fig.~\ref{fig:visib}, right column, 3rd panel from the bottom,  which show higher dispersion than error bars predict as well as some bias towards a less resolved star).  GJ~785 is a K star with no hint of specially high activity, so it is unlikely that this $V^2$ dispersion is produced by a short-term variability.  A relatively fast atmospheric turbulence ($\tau_0 = 5$--7\,ms in $H$ with a decent seeing of 0.7--0.9) may explain part of it. Two other stars of the sample show a high, unexplained dispersion of measurements at several baselines which could be imputed to an intrinsic short-term variability: for M-type flare stars GJ~54.1 and GJ~447, $\sigmauncorr=4.3$\% and 6.1\%, respectively but, in their case, the dispersion only impacts the uncertainty, not the diameter estimate.  For the remaining 85\% of the sample, the unexplained dispersion is moderate ($\sigmauncorr = 2.3\pm0.9$\%) and the \modelcov{} \& \modelcovbl{} diameter estimates are consistent with each other.

\subsection{Systematic errors}
\label{sec:influence-syserror}
\begin{figure}
\includegraphics[width=\linewidth]{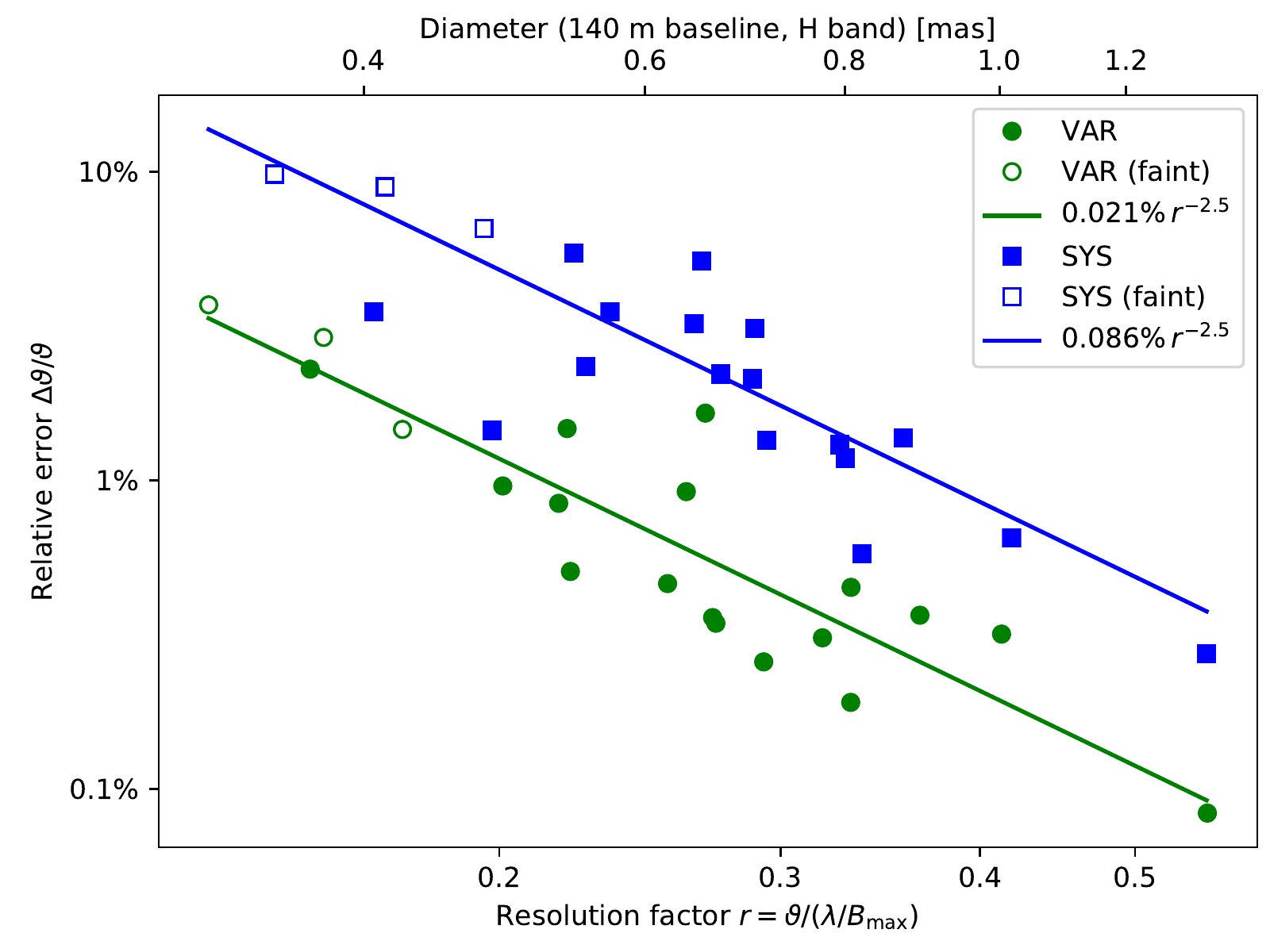}
\caption{Relative uncertainty on the diameter versus the resolution factor $r = \vartheta/(\lambda/B)$, with the standard processing technique (\modelvar{} model, green) and our determination with covariances and systematics (\modelsys{} model, Sect.~\ref{sec:influence-syserror}). Stars observed close to, or below, the nominal sensitivity limit of PIONIER are represented as hollow markers.}
\label{fig:delta-ud}
\end{figure}

Usually, error models assume that the uncertainties on data points are underestimated and rescale them so that a reduced chi squared of 1 is obtained. This is indeed what we do in our models that exclude systematic errors (\modelvar{} and \modelcov{} model groups). 

In the \modelsys{} model group, however, no rescaling of error bars occurs. Instead a relative error term $\sigmasys$ is introduced, as explained in Sect.~\ref{sec:syserr} and the fit is performed with a correlated covariance matrix ($\ubcovar$ in Eq.~\ref{eq:syscovar}).  The value of $\sigmasys$ is fit to a reduced chi squared of one.  

The difference in estimated diameter is often significant as 25\% of the stars show a discrepancy of at least 1-$\sigma$ (see e.g.\ Fig.~\ref{fig:estim-comp}, bottom-left panel) between the \modelcov{} and \modelsys{} models.

In Fig.~\ref{fig:ud-comp}, one can also see (left panel, 6th and 9th points from the left, labelled \modelcovbl{} and \modelsys) that systematics tend to increase the diameter uncertainties (as positive correlation usually do).  A comparison of uncertainties in the \modelvar{} and \modelsys{} models is given in Fig.~\ref{fig:delta-ud} as a function of the resolution factor, which we define as the ratio of the UD diameter of the star to the nominal resolution of the interferometer $\lambda/B_\mathrm{max}$.   The typical error in \modelsys{} is 4.8\% for a resolution factor of 0.2 ($\approx 0.45$\,mas diameter at 140\,m baselines in the $H$ band) with respect to 1.2\% in the standard \modelvar{} model.  Since the \modelvar{} model is widely used in the literature, we conclude that diameter uncertainties may be significantly underestimated. The relative uncertainty scales approximately as the inverse of the 2.5th power of the resolution, making determinations for under-resolved objects ($\lesssim 0.4$\,mas at VLTI in $H$) difficult.  Since the fringe contrast loss $1-V^2$  scales as the square of the diameter, we expected the diameter uncertainty to scale as the inverse of the second power of the resolution factor. The additional drop in precision may be attributed to a loss of precision as the stars get fainter.  In Fig.~\ref{fig:ud-comp}, we can see that 3 of the 4 most under-resolved targets have been observed very close to the sensitivity limit of PIONIER.

For reference, Fig.~\ref{fig:ud-comp} shows another error model, \modelppp, that uses the na\"\i{}ve but erroneous correlation matrix leading to Peelle's pertinent puzzle.  Indeed, the diameter estimates are significantly off for about half of the sample.

\subsection{Noise in the covariance matrix}
\label{sec:influence-noise}

With model \modelvarnoisy{} we aim to assess the influence of the noise in the correlation to disentangle noise systematics from actual correlation influences. We measure the noise level in the covariances $\covar$, generate noisy matrices for uncorrelated data, and perform a data fit using these matrices.
 
Most data are only lightly correlated, for different nights and instrument configurations are loosely correlated by the small uncertainty on the calibrators diameter. So, the histogram of the correlation matrix values features a central component, a Gaussian-like distribution around a small positive constant, corresponding to the mostly uncorrelated data. In addition it has a bump with large positive correlations corresponding to the small fraction of highly correlated data (spectral channels of the same observation, for instance).  We fit the width of the central component to get the noise level.

Then, we generate $\Nboot$ covariance matrices for noisy uncorrelated data
\begin{subequations}
\begin{align}
    \correl           &= \delta_{\obsi\obsj} 
                + \varepsilon \sum_k M_{k\obsi}^{(\booti)} M_{k\obsj}^{(\booti)}\\
    \noisycovar^{(\booti)} &= \correl \dcalvisi \dcalvisj
       \label{eq:noisycovar}
\end{align}
\end{subequations}
where $M_{k\obsi}^{(\booti)}$ are picked using independent normal Gaussian distributions and $\varepsilon$ is adjusted to the noise level.  The largest dimension along $k$ that allows \smash{$\correl$} to be positive definite is used. 

For each of these noisy covariance matrices, we perform a least-squares fit to all calibrated visibilities $\calvis{}$ taken at baselines $\base$.  Let \smash{$\diamnoisy$}  ($\booti$ in $1 \cdots \Nboot$) be the values of the UD diameters. The UD diameter estimate $\diamnoisy[]$ is obtained from their average. Its uncertainty $\ddiamnoisy$ is the quadratic sum of the model uncertainty $\ddiamvar$ and the dispersion of \smash{$\diamnoisy$}.

As we can see from the data in Fig.~\ref{fig:ud-comp} (left panels, second and fourth point), there is no significant difference between the \modelvar{} and \modelvarnoisy{} models for any of our stars.  We conclude that the impact of correlations is a real effect, not a bias introduced by noisy data.

\subsection{Non-Gaussian uncertainties}
\label{sec:influence-gauss}

In order to assess the influence of non Gaussian uncertainties, we performed least squares fit to each bootstrap and look at the distribution of the estimates for the uniform disc diameter.  We expect that a distribution of visibilities with significant skewness, kurtosis, or long tails would be reflected in the distribution of diameter estimates, and, in turn, in its mean value and/or uncertainty.
 
For model \modelvarboot, $\calvisboot \pm \dcalvis$ ($1 \le b \le \Nboot$) is fitted, yielding a set of uniform disc diameter estimates $\diamvarboot$. The median value and 1-$\sigma$ confidence interval of $\diamvarboot$ yields the \modelvarboot{} diameter estimate $\diamvar \pm \ddiamvar$.  The same is performed for error models with correlated statistical errors (\modelcovboot) and correlated statistical and systematic errors (\modelsysboot).  

Figure~\ref{fig:ud-comp} (left panel) clearly shows that there is no significant difference in the diameter estimate for \modelvar{} and \modelvarboot, \modelcovbl{} and \modelcovboot, and \modelsys{} and \modelsysboot.   
 
\subsection{``Bad baseline'' scenario}
\label{sec:influence-badbase}

We additionally check for the scenario that a ``bad baseline'', e.g. tainted with a strong systematic, may significantly alter the value of the uniform disc estimate. Given that baseline, we fit the data that are taken at any other baseline and compare the diameter estimates with the one obtained with all baselines.  The process is repeated for each baseline.  The diameter estimates with one baseline removed are reported on the right side of Fig.~\ref{fig:ud-comp}. 

\begin{figure*}
\centering
\includegraphics{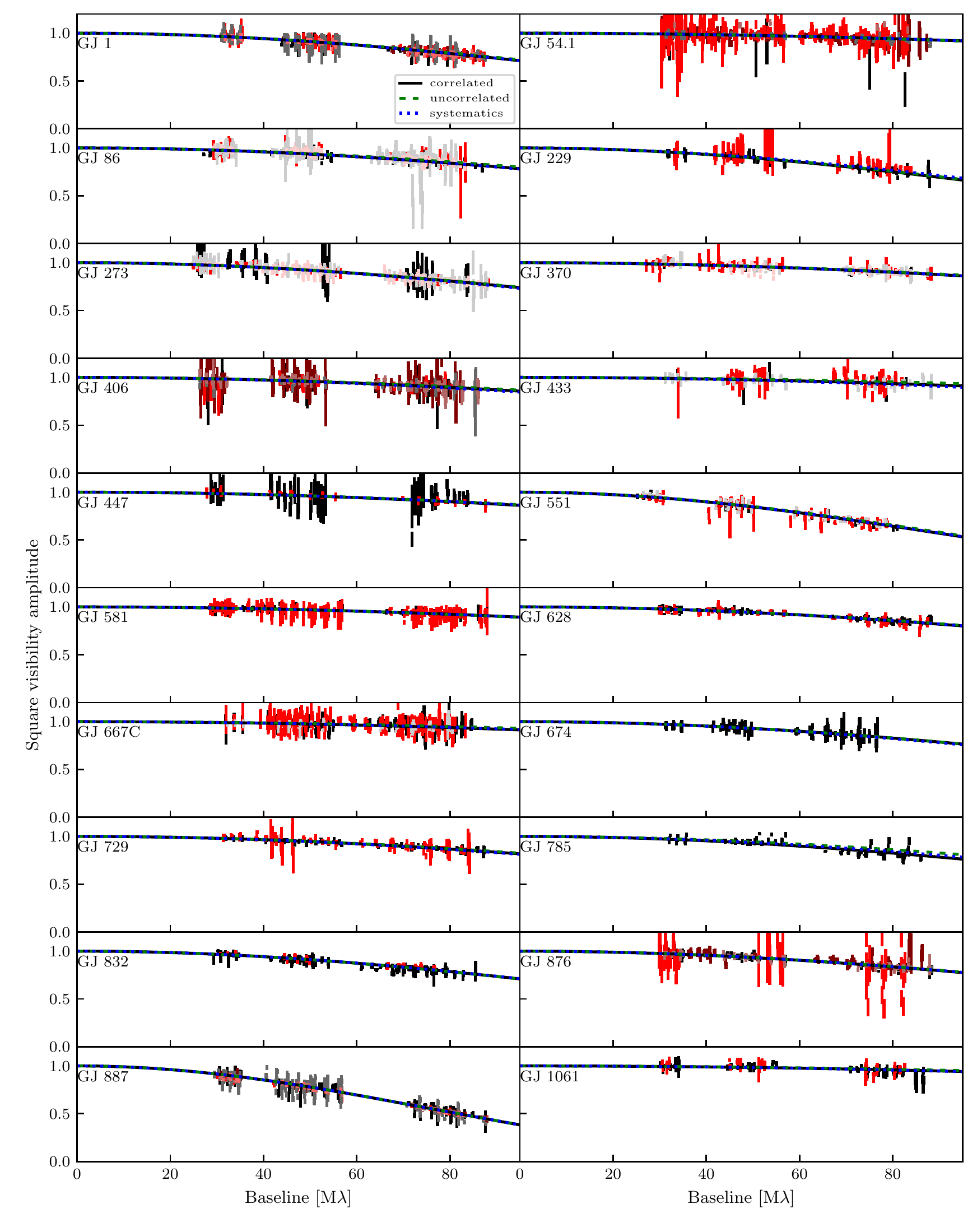}
\caption{Squared visibility amplitudes versus baseline length. \textit{Vertical error bars:} reduced data, using one shade of grey and red per baseline. \textit{Lines:} best model fits \modelvar{} (uncorrelated, green dashed line), \modelcovbl{} (correlated, black solid line), \modelsys{} (with systematics blue dotted line). In many models they can hardly be distinguished from one another.} 
\label{fig:visib}
\end{figure*}

Two examples o. ``bad baselines'' can be seen in Fig.~\ref{fig:visib}.   For GJ~876, a baseline around 75 to 80 megacycles displays very high dispersion ($V^2$ between 0.5 and 1.2) despite small error bars ($\sim 0.05$). This issue does not originate from a bad calibrator, because this kind of dispersion would only be seen with a well resolved, low contrast binary. That would clearly been seen (1) in closure phases which was not the case and (2) as a strong bias towards a more resolved object with most points below the fit.   The inclement weather during this particular observation is a likely explanation with a seeing below average (1.05 arcsec in $H$) and a short coherence time (6 ms in $H$). Interestingly, this ``bad baseline'' has very little impact on the diameter estimates as GJ~876 has all its estimates (\modelvar, \modelcov, \modelcovbl, \modelsys{} models) within less than 1\% \change{dispersion}, probably because it shows little bias (the data approximately average the $V^2 \approx 0.85$ estimate).  Another example of ``bad baseline'' has already been mentioned for GJ~785 (less resolved around 50 and 70-80\,M$\lambda${} in Fig.~\ref{fig:visib}, left column, 3rd panel from the bottom), this time with a clear bias. Its origin is unknown, but it is unlikely to be a calibrator issue either: if there was a large error on the diameter of the unique, large calibrator HD~196387, there would be an unseen error on GJ~785's diameter but very little additional dispersion between baselines.  The calibrator has no measured closure phase ($\lesssim 0.02$\,rad) and no detectable near-infrared excess \citep[no evidence from JHK modelling in][]{MCD12}, so binarity or circumstellar material are very unlikely to account for a 5\% error in the visibility calibration. The bias induced by the ``bad baseline'' reflects in Fig.~\ref{fig:ud-comp}, right panel: with this baseline removed (3rd point of the 6), the estimates of the \modelvar, \modelcovbl, and \modelsys{} models are consistent, but with it (left panel) they differ by several $\sigma$.

These two ``bad baselines'' are extreme cases that could be solved easily by removing the data from the fit. However, there are targets (GJ~86, GJ~229,  GJ~447, GJ~551) where the visibility plot (Fig.~\ref{fig:visib}) does not show anything obviously wrong, but the analysis of removing one baseline in the fits produces a significant difference in the \modelvar{} and/or \modelcovbl{} diameter estimates (right panel of Fig.~\ref{fig:ud-comp}).  

When systematic errors are included, the impact of ``bad baselines'' disappears completely (right panel of Fig.~\ref{fig:ud-comp}): all diameter fits with one baseline removed are consistent with each other.  The price to pay for this stability is larger error bars as we have shown in Sect.~\ref{sec:influence-syserror} and Fig.~\ref{fig:delta-ud}.

It appears, therefore, that observations with a few configurations over a few nights are enough to prevent a single baseline to significantly bias the diameter estimate as long as systematic errors are modelled.

\subsection{Bad calibrator scenario}

For several of our targets (GJ 1, GJ 54.1, GJ 86, GJ 370) one or more calibrators from MER05 more resolved than the target were used together to smaller, unresolved ones. In the bad luck scenario that a resolved calibrator's diameter is off by several standard deviations, it would bias the transfer function at the two or three large baselines it is observed with. Fortunately, for these four stars, there is no evidence that a few baselines skew the diameter estimate. The difference in the COV estimates when one baseline removed from the fit is $< 1$\% (right panel of Fig.~\ref{fig:ud-comp}) and the COV and SYS estimate are within 1\% of each other.  We are confident that the diameter estimates of these stars are not significantly impacted by a bad calibrator.

\section{Conclusion}
\label{sec:conclusion}

Many astronomical objects (young stellar objects, late-type dwarfs) are too faint in the visible to be observed by optical interferometers and require kilometric baselines to be fully resolved in the infrared.  For this reason, we are bound to infer the object's properties from under-resolved observations.   In the case of stellar diameters, we are fortunate enough that there are no degeneracies in the (unique) parameter estimation, but the under-resolved character has a strong impact in terms of precision.  While the geometric size of a fully resolved object can be estimated within a fraction of a percent, uncertainties and systematics of 5--10\% on a single observation are common when the target is under-resolved.

Our study had the main objective to partially overcome these limitations by a careful observation layout and to better quantify the remaining uncertainties. To that end, our \modelsys{} model fully takes into account correlated statistical uncertainties and systematics.

The main results are
\begin{itemize}
    \item The error model has a significant impact on the estimate of the uniform disc diameter and its uncertainty.  Correlations between visibilities have a strong impact (Sects.~\ref{sec:influence-corr}) as well as baseline systematics (Sect.~\ref{sec:influence-syserror}).  In both cases, the estimates may differ by more than three standard deviations. The additional errors atop the statistical errors determined by the data processing software are $3.3 \pm 1.2$\,\% on the square visibilities (mean and dispersion in our 20 surveyed stars), which is in line with the generally accepted value of 5\% \citep{COL03}. However, the purely systematic term (highly correlated errors) is only $1.8 \pm 0.9$\,\%.
    \item A few observations with different configurations and/or nights are usually enough to avoid a significant bias by a single baseline and instrumental configuration (Sect.~\ref{sec:influence-badbase}) provided that systematic errors are taken into account. It confirms the usual observation strategy by, for instance, \citet{BOY12}, \citet{GAL12}, and \citet{RAB18} in the case of stellar diameters of under-resolved stars.
    \item Departure from a Gaussian distribution has no significant impact (Sect.~\ref{sec:influence-gauss}) except indirectly in the error propagation (Sect.~\ref{sec:influence-prop}).  It needs not to be modelled provided that the uncertainties on the calibrated visibilities are obtained from the bootstraps (this work) or from an analytic determination of the probability density function \citep{PER03}.
    \item The uncertainty on the diameter is four times larger than the ones modelled with a standard least-squares fit to uncorrelated data.  For a diameter of 0.45\,mas in $H$ band at 140\,m baselines (typical of VLTI/PIONIER), our typical uncertainty is 4.8\% (\modelsys{} model) with respect to 1.2\% with the usual determination (\modelpndrs{} and \modelvar{} models).  
\end{itemize}

We have offered here a relatively easy way, albeit numerically intensive, to obtain correlations between observables, by means of bootstrapping.  Even if most reduction pipelines do not propagate correlations, it is possible to run them a large number of times on randomised data (interferograms and calibrator diameters are picked at random) to obtain the multivariate probability density function of the interferometric observables. Systematics errors such as a bad calibrator not bad enough to be detected, rapidly varying atmospheric conditions between science target and calibrators, or instrumental systematics (e.g. differential polarisation) have typically cast a doubt on the robustness of parameter estimation.  We have provided a method to deal with these systematics in a relatively inexpensive way: the covariance matrix of the least-squares fit is modified to include a relative systematic error term. The price to pay for a more robust estimation (accurate, i.e. non-biased) is a significantly larger uncertainty.

Therefore, we strongly recommend that future interferometric studies take into account correlated errors and take time to model systematics.

\section*{Acknowledgements}

Based on observations collected at the European Organisation for Astronomical Research in the Southern Hemisphere under ESO IDs 090.D-0917, 091.D-0584, 092.D-0647, and 093.D-0471. R.L., M.R., A.J., and R.B. acknowledge support from CONICYT project Basal AFB-170002.   A.J. acknowledges support from FONDECYT project 1171208, BASAL CATA PFB-06, and project IC120009 ``Millennium Institute of Astrophysics (MAS)'' of the Millenium Science Initiative, Chilean Ministry of Economy. R.B. acknowledges additional support from project IC120009 ``Millenium Institute of Astrophysics (MAS)'' of the Millennium Science Initiative, Chilean Ministry of Economy. This work made use of the Smithsonian/NASA Astrophysics Data System (ADS) and of the Centre de Données astronomiques de Strasbourg (CDS). This research made use of Astropy, a community-developed core Python package for Astronomy (Price-Whelan et al. 2018).  We would like to thank the referee for carefully reading our manuscript and for giving constructive comments which substantially helped improving the paper.

\bibliographystyle{mnras}
\bibliography{bootstrap}

\end{document}